\pdfoutput=1
\documentclass[a4paper,11pt]{article}
\usepackage{amssymb,lscape,psfrag}
\usepackage{epsfig,color}
\usepackage{array}
\usepackage{enumerate}
\usepackage{amsmath}
\usepackage{graphics}
\usepackage{graphicx}
\usepackage{epsfig}
\usepackage{verbatim}
\usepackage{cancel}
\usepackage{longtable}
\usepackage{multirow}
\usepackage{float}
\usepackage[normalem]{ulem} 
\usepackage[toc,page]{appendix}
\usepackage{rotating}
\usepackage{amsmath}
\usepackage{url}
\usepackage{rotating}
\usepackage{hyperref}
\usepackage{mathabx}
\usepackage{xcolor}
\usepackage{natbib}
\hypersetup{
	colorlinks,
	linkcolor={red!75!black},
	citecolor={blue!50!black},
	urlcolor={black!100!black}
}

\usepackage{natbib}

\newcommand {\inispace}{\renewcommand{\baselinestretch}{1.}\normalsize}

\newcommand {\bmath} {\begin {displaymath} }
\newcommand {\emath} {\end {displaymath} }

\newcommand {\by} {\mathbf{y}}
\newcommand {\diag} { \mbox{diag} }
\newcommand{\bbhat}{\boldsymbol{\widehat{\beta}}}
\newcommand{\ghat}{\boldsymbol{\widehat{\gamma}}}
\newcommand{\bb}{\boldsymbol{\beta}}
\newcommand{\g}{\boldsymbol{\gamma}}
\newcommand{\bbj}{\boldsymbol{\beta}_k}
\newcommand{\I}{\boldsymbol{I}_n}
\newcommand{\Ip}{\boldsymbol{I}_{p}}
\newcommand{\Ij}{\boldsymbol{I}_{p_k}}
\newcommand{\X}{\mathbf{X}}
\newcommand{\XT}{\mathbf{X}^{T}}
\newcommand{\Xj}{\mathbf{X}_{k}}
\newcommand{\XTj}{\mathbf{X}_{k}^{T}}
\newcommand{\G}{\mathbf{G}}
\newcommand{\Gl}{\mathbf{G}_{\boldsymbol{\lambda}}}
\newcommand{\wl}{\mathbf{w}_{\boldsymbol{\lambda}}}
\newcommand{\Hl}{\mathbf{H}_{\boldsymbol{\lambda}}}

\newcommand{\La}{\mathbf{\Lambda}}
\newcommand{\Laj}{\mathbf{\Lambda}_{k}}

\newcommand{\N}{\mathrm{N}}
\newcommand{\CV}{\mathrm{CV}}
\newcommand{\ML}{\mathrm{ML}}
\newcommand{\PM}{\mathrm{PM}}
\newcommand{\bl}{\boldsymbol{\lambda}}
\newcommand{\Sb}{\boldsymbol{\Sigma}_{\boldsymbol{\beta}}}
\newcommand{\Vb}{\mathbf{V}_{\boldsymbol{\beta}}}
\newcommand{\xii}{\mathbf{x}_{i}}
\newcommand{\xti}{\mathbf{x}_{i}^T}
\newcommand{\A}{\mathbf{A}}
\newcommand{\Ainv}{\mathbf{A}^{-1}}
\newcommand{\CL}{\textnormal{\tiny \textsc{CL}}}
\newcommand{\RNA}{\textnormal{\tiny \textsc{RNA}}}
\newcommand{\SNP}{\textnormal{\tiny \textsc{SNP}}}
\newcommand{\M}{\mathbf{M}}
\newcommand{\U}{\mathbf{U}}
\newcommand{\D}{\mathbf{D}}
\newcommand{\V}{\mathbf{V}}

\DeclareMathOperator*{\argmin}{argmin}
\DeclareMathOperator*{\argmax}{argmax}

\begin{document}

\inispace
\title{Scalable Bayesian regression in high dimensions \\ with multiple data sources}
\date{}
\author{Konstantinos Perrakis$^{\dagger}$, Sach Mukherjee$^{\dagger}$\\
	and\\
	The Alzheimers Disease Neuroimaging Initiative\footnotemark
	\\[0.3cm]
	{\it \small $^{\dagger}$Statistics and Machine Learning,} \\ {\it  \small  German Center for Neurodegenerative Diseases (DZNE),  Bonn, Germany.} 
}

\maketitle

{\let\thefootnote\relax\footnotetext{*Data used in preparation of this article were obtained from the Alzheimers Disease Neuroimaging Initiative (ADNI) database (adni.loni.usc.edu).  As such, the investigators within 	the ADNI contributed to the design and implementation of ADNI and/or provided data but did not participate in analysis or writing of this report.  A complete listing of ADNI investigators can be found at:
		\url{http://adni.loni.usc.edu/wp-content/uploads/how_to_apply/ADNI_Acknowledgement_List.pdf}}}

\pagenumbering{arabic}

\begin{abstract}
Applications of high-dimensional regression often involve multiple sources or types of covariates. We propose methodology for this setting, emphasizing  the ``wide data" regime with   large total dimensionality $p$ and sample size $n {\ll}  p$. We focus on a flexible ridge-type prior with shrinkage levels that are specific to each data type or source and that are set automatically by empirical Bayes. All estimation, including setting of shrinkage levels, is formulated mainly in terms of inner product matrices of size $n {\times}  n$. This renders computation efficient in the  wide data regime and allows scaling to problems with millions of features. Furthermore,  the  proposed procedures  are free of user-set tuning parameters. We show how sparsity can be achieved  by post-processing of the Bayesian output via constrained minimization of a certain Kullback-Leibler divergence. This yields sparse solutions with adaptive, source-specific shrinkage, including a closed-form  variant that scales to very large $p$. We present empirical results from a simulation study based on real data and  a  case study in  Alzheimer's disease involving millions of features and multiple data sources.
	
	\vspace{0.2cm} \noindent \textit{Keywords: Bayesian post-processing, high-dimensional regression, multiple data types, prediction, ridge regularization, shrinkage priors} 
	
\end{abstract}

\section{Introduction}

	Advances in data acquisition have meant that studies in many fields now
routinely include multiple sources of features, such as  different data types, with one or more of the sources being high-dimensional. 
To fix ideas, consider a biomedical setting in which samples indexed by $i \! = \! 1, \ldots, n$ each have response $y_i$ and features of several types  $k \! = \! 1, \ldots, K$ (representing say genetic data, imaging, clinical covariates and so on)  with respective dimensionalities $p_1, \ldots ,p_K$.
We refer to the different types of feature as sources. The $p_k$'s are the source-specific dimensionalities and 
$p \! = \! \sum_{k=1}^K p_k$ is the total dimensionality. 
We consider  a specific example of this kind below, in the context of Alzheimer's disease.

Constructing regression models using such data is challenging, because the relevance of the sources may be quite unequal (and unknown in advance) and the total dimension $p$ may be  large. This motivates a need for methodology that can 
cope with multiple sources and that scales to  high dimensions. 

Methods for high-dimensional regression are now well established and include 
regularized least-squares approaches such as the lasso and extensions \citep{tibshirani,tibshirani2005,Yuan_Lin}, elastic net \citep{zou_hastie_2005}, SCAD (\citealp{fan_li_2001}), and  Bayesian analogues \citep[see][for a review]{Kyung}.
A range of Bayesian approaches have been proposed, notably those based on shrinkage priors, often coupled with variable selection, see for instance  \cite{Yuan_Lin_2005,park_casella,hans_2010,griffin_brown_2010,carvalho_etal_2010}, and \cite{armagan_etal_2012}, among  others.
However, in the very large $p$ case many available methods become computationally cumbersome or intractable  and effective source-specific shrinkage remains hard to achieve.

In this paper we put forward an approach to regression in the multiple-source, high-dimensional  setting. Specifically:
\vspace{-0.1cm}
\begin{itemize}
	\item We consider  a generalized ridge-type prior with shrinkage  that adjusts to individual data sources, with the corresponding shrinkage levels  estimated from the data.
	\vspace{-0.2cm}
	\item We show that estimation (including setting of tuning parameters)
	can be formulated in a way that renders computation  efficient for ``wide" data, even for very large $p$ and over multiple sources.
	
	\vspace{-0.2cm}
	\item We introduce sparsifications that achieve competitive prediction performance and that provide a fast yet multivariate technique for discarding non-influential features.
\end{itemize}
Thus, we consider the case of data from multiple sources with source-specific dimensionalities $p_k$ that could differ by many orders of magnitude, with total $p$  large and {\it a priori} unknown source-specific importance. The main strength of our methods is  their ability to use source-specific shrinkage to automatically adapt to signals spread across multiple sources. 


There has been much interesting work on group selection approaches in regression \citep[reviewed in][]{huang2012}. The group lasso \citep{Yuan_Lin} allows specification of covariate subsets that can then be selected as groups; 
however, applying the group lasso in the current setting (by identifying groups with sources) would not be useful because sources would then simply be either included or excluded (without within-source regularization). 
The sparse group lasso \citep{simon2013} permits additional regularization via within-group sparsity but its use here would require 
a nontrivial  extension to source-specific penalties whose tuning would be difficult if not  intractable in the very high-dimensional, multiple source setting.
\citet{Dondelinger_Mukherjee} consider the case of penalized regression over multiple subgroups of samples; this is 
quite different from the present setting of sources of covariates (i.e., we focus on the columns, not the rows), also the authors do not tackle the very high-dimensional case.

Ridge-type estimators are among the oldest and best studied regularized regression tools, whether from a penalized likelihood or Bayesian viewpoint. Our results build on these classical tools, developing a variant of the ridge prior to deal with multiple-source, high-dimensional problems.
The subsequent sparsification step that we consider is an example of an emerging class of posterior post-processing methods and yields a solution which is similar to the penalized credible region \citep[pCR,][]{bondell_reich2012} and the decoupled shrinkage selection \cite[DSS,][]{hahn_carvalho2015} solutions. In contrast to pCR and DSS 
we develop our approach via a certain Kullback-Leibler divergence.
Interestingly, we can recover the pCR solution as a special case.
Our approach can take advantage of the information from the initial ridge step and thus allows parameter-specific as well as source-specific penalization.
In addition, we propose a relaxed variant which leads to a closed-form solution that is immediately applicable to problems involving millions of predictors.
The primary motivation for this work is the need for efficient and interpretable predictive models in high-dimensional biomedical applications. We emphasize that the sparse extensions proposed are mainly aimed at achieving parsimonious prediction
rather than variable selection {\it per se}. However, we also discuss some preliminary empirical results on variable selection using the class of methods proposed here.


As a topical example of a multiple-source, high-dimensional  problem, 
we consider a case study in Alzheimer's disease (AD).
AD is a neurodegenerative condition in which 
prediction of future disease course is a central research topic.
AD is 
multifactorial in the sense of being mediated via multiple underlying biological processes and 
several current and emerging large-scale studies span multiple data types. These include the
Alzheimer's Disease Neuroimaging Initiative (ADNI) \citep{ADNI1}, the Rhineland study (\url{http://www.rheinland-studie.de}), 
and the UK Biobank (\url{http://www.ukbiobank.ac.uk}) (this is a broader study including neurodegeneration-related data).
The data we consider are from ADNI which is 
a large-scale longitudinal study involving multiple data modalities; we focus specifically 
on the prediction of future cognitive scores, as described in detail below.

The remainder of this paper is organized as follows. In Section \ref{SBR} we introduce the scalable Bayesian regression (SBR) approach, describing model formulation, prior specification, and tuning of shrinkage levels. Section \ref{Sec:SSBR} deals with the sparse extension of the methodology, sparse SBR (SSBR), including  a general solution and a relaxed variant for the very large $p$ case. The relationship between SSBR and pCR is discussed. We further introduce an adaptive approach to regulate induced sparsity.
Results and comparisons with standard penalized likelihood approaches from a simulation study are presented in Section \ref{SIM_STUDY}, while the AD case study appears in Section \ref{ADNI}.
The paper concludes with a discussion
in Section \ref{Disc}.

\section{Scalable Bayesian regression}
\label{SBR}


\subsection{Model}
\label{likelihood}
Let $\by$  be a $n \times 1$ vector of  responses  and $ \X_1, \dots, \X_K $ denote covariate or feature matrices from $K$ data sources. Each $\Xj$ is of size $n\times p_k$ so that the total number of potential predictors is $p=\sum_{k=1}^{K}p_k$. We consider the normal linear model 
\begin{equation*}
\by = \X_1\bb_1 + \X_2\bb_2 + \cdots + \X_K\bb_K + \boldsymbol\varepsilon, \quad \boldsymbol{\varepsilon}\sim \N_n(\mathbf{0},\I\sigma^2), 
\end{equation*}
where 
each $\bb_k$ is a $p_k$-vector of regression coefficients, $\N_n(\mu,\Sigma)$ denotes an $n$-dimensional multivariate normal density with mean $\mu$ and covariance $\Sigma$, and $\boldsymbol{I}_n$ is the $n\times n$ identity matrix. 
Without loss of generality we assume throughout that all data are standardized. Let $\X = [\X_1 \cdots \X_K]$ and $\bb = (\bb_1^T, \cdots, \bb_K^T)^T$ denote the respective global $n\times p$ predictor matrix and $p$-vector of  regression coefficients (here, ``global" means with all sources taken together). 
Then, with prior $\pi$ the full model under consideration is 
\begin{equation*}
\by \sim  \N_n(\X\bb,\I\sigma^2), \, \, \mbox{with    } \bb|\sigma^2 \sim  \pi(\bb|\sigma^2) \mbox{    and    } \pi(\sigma^2) \propto 1/\sigma^{2}.
\end{equation*}
The improper prior for $\sigma^2$ (Jeffreys' prior) is a common option for linear regression models. The crucial aspect of prior formulation for the multiple-source, high-dimensional setting under consideration is the construction of $\pi(\bb|\sigma^2)$, as we discuss in detail below.

\subsection{The prior on $\bb$}
\label{Sec:Lambda}

The SBR approach is based on a  natural generalization of the standard ridge prior. 
Specifically, the prior on $\bb$ is
\begin{equation}
\bb \mid \bl,\sigma^2 \sim \N_p(\mathbf{0},\La^{-1}\sigma^2),
\label{Lambda_prior}
\end{equation}
where $\bl=(\lambda_1,\cdots,\lambda_K)^T$, $\La=\diag(\La_1,\cdots,\La_K)$, and $\Laj=\lambda_k\Ij$ with $\lambda_k>0$, for $k=1, \dots, K$. Here each $\lambda_k$  is a source-specific shrinkage level on the corresponding $\bb_k$. The special case  $K{=}1$ recovers the standard ridge prior with just one shrinkage level (and indeed the solutions presented below
could be used to give a scalable implementation of classical ridge with a single $\lambda$).
However, when dealing with multiple data sources the various data sources may differ in importance. This motivates a need for source-specific  penalties that can adjust to account for such differences and additionally  provide potentially useful information about the relevance of specific data sources.

At this point it is useful to define the  quantity
\begin{equation}
\G_{\boldsymbol{\lambda}} \overset{\mathrm{def}} = \sum_{k=1}^{K}\lambda_k^{-1}\Xj\XTj.
\label{Gram}
\end{equation}     
All formulas presented in the remainder of this Section are cast in terms of $\G_{\boldsymbol{\lambda}}$. Importantly, this means that the key computations under SBR can be formulated so as to require only a one-off computation of these individual inner product (Gram) matrices $\Xj\XTj$ of size $n \times n$ (these calculations can be easily implemented in parallel) followed mainly by operations on those matrices. As we show below, for  wide data with  large $p$, this gives a practical way to implement SBR.

\subsection{Inference}

Under the conjugate prior in \eqref{Lambda_prior} the posterior distribution of $\bb$ is given by 
\begin{equation}
\bb \mid \by,\bl,\sigma^2 \sim \N_p(\bbhat,\Sb\sigma^2),
\label{Lambda_post}
\end{equation} 
where $\bbhat=\Sb\XT\by$ and  $\Sb=(\XT\X+\La)^{-1}$.
Calculating the posterior mode directly involves a $p\times p$ matrix inversion. For  $p>n$ we instead use
\begin{equation}
\bbhat=\La^{-1}\XT\wl,
\label{Lambda_mode}
\end{equation}
where
$\wl=\big[\by-(\I+\Gl)^{-1}\Gl\by\big]$
is an $n$-vector whose calculation involves only an $n \times n$ matrix inversion. The derivation of \eqref{Lambda_mode} is provided in Appendix \ref{AppA} of the supplementary material. For very large problems the computation of the posterior mode can be done in parallel; additionally, we draw attention to the useful expression
\begin{equation}
\bbhat_k=\lambda_k^{-1}\XT_k\wl,
\label{Lambda_mode1}
\end{equation} 
for $k=1,\dots, K$.
Having obtained the posterior mode, prediction from an available $\X^{\mathrm{pred}}$ of dimensionality $m\times p$ is straightforward via $\by^{\mathrm{pred}}=\X^{\mathrm{pred}}\bbhat$. When interest lies solely in prediction the corresponding calculation can be simplified to
\begin{equation}
\by^{\mathrm{pred}}  =\X^{\mathrm{pred}}\La^{-1}\XT \wl\\
=\Bigg[\sum_{k=1}^{K}\lambda_k^{-1}\Xj^{\mathrm{pred}}\XTj\Bigg] \wl.
\label{Lambda_pred}
\end{equation}
Calculating the posterior covariance matrix can also be simplified through the formula
\begin{equation}
\Sb =  \La^{-1}\big[\Ip-\XT(\I+\Gl)^{-1}\X\La^{-1}\big].
\label{Cov}
\end{equation}
For details see Appendix \ref{AppA}. 

In practice we are not interested in evaluating the entire covariance matrix (for very large $p$ this might in fact be difficult due to memory limitations). However, the methodology considered in Section \ref{Sec: Rel_sol} does require the diagonal elements of $\Sb$. In this case the formula in \eqref{Cov} facilitates computation as it allows for fast and parallel block matrix computations. Note that 
$  \boldsymbol{\Sigma}_{\bbj}= \lambda_k^{-1}\big[\boldsymbol{I}_{p_k}-\XT_k(\I+\Gl)^{-1}\X_k\lambda_k^{-1}\big]$, so here the magnitude of each $p_k$, for $k=1,\dots,K$, can guide us in determining whether to use block computations or not. 
To clarify the use of parallel computations here, consider calculating the diagonal elements of a specific large $\boldsymbol{\Sigma}_{\bbj}$ using $B$ blocks. Consider $\X_k=[\X_k^{(1)} \cdots \X_k^{(B)}]$ where each $\X_k^{(b)}$ has $p_k^{(b)}=\left\lfloor p_k/B \right\rfloor $ columns for $b=1,\cdots, B-1$, while $p_k^{(B)}=p_k-(B-1)p_k^{(1)}$ for the last sub-matrix $\X_k^{(B)}$. Denote the variances by $\boldsymbol{\sigma}_{\bb_k}^{2}=\diag\big(\boldsymbol{\Sigma}_{\bbj}\big)$. The calculation which is then performed in parallel is
\begin{equation}
\boldsymbol{\sigma}_{\bb_k}^{2(b)}=\diag\Big(\lambda_k^{-1}\Big[\boldsymbol{I}_{p_k^{(b)}}-\X_k^{(b)T}(\I+\Gl)^{-1}\X_k^{(b)}\lambda_k^{-1}\Big]\Big),
\label{parallel}
\end{equation}
for $b=1,\cdots,B$, with the output being the $p_k$-dimensional vector of variances; namely, $\boldsymbol{\sigma}_{\bb_k}^{2}=(\boldsymbol{\sigma}_{\bb_k}^{2(1)},\cdots,\boldsymbol{\sigma}_{\bb_k}^{2(B)})^T$.

Continuing, the posterior distribution of the error-variance parameter is an inverse-gamma distribution; specifically 
\begin{equation}
\sigma^2 \mid \by\sim\mathrm{IG}(a,b), \mbox{~with shape~} a=\frac{n}{2} \mbox{~and scale~} b=\frac{\by^T(\I+\Gl)^{-1}\by}{2}.
\label{post_sigma}
\end{equation}
The numerator of the scale parameter in \eqref{post_sigma}, whose calculation is again simplified via $\Gl$, is important because it will appear throughout; for its derivation see Appendix \ref{AppA} in supplementary material.

Below we will make use of the marginal likelihood $m(\by|\bl)$. This is the likelihood obtained by integrating over the parameter space with respect to the joint prior distribution of $\bb$ and $\sigma^2$.
Under our conjugate prior specification we have 
\begin{equation}
m(\by|\bl) \propto \int p(\by|\bb,\sigma^2) \, \pi(\bb|\bl,\sigma^2) \, \frac{1}{\sigma^2} \, \mathrm{d}\bb \, \mathrm{d}\sigma^2 \propto |\I+\Gl|^{-\frac{1}{2}}\big[\by^T(\I+\Gl)^{-1}\by\big]^{-\frac{n}{2}} \, .
\label{marg_lik}
\end{equation}

\subsection{Automatic setting of shrinkage levels $\bl$}
\label{tuning}

Specification of penalty parameters is often handled through cross-validation (CV) or generalized CV in a frequentist framework \citep{tibshirani}, while Bayesian methods typically rely on empirical Bayes (EB) point estimates or data-dependent hyper-priors; see e.g.,  \cite{Yuan_Lin_2005,park_casella,bala_madigan_2010,hans_2010}, and \cite{griffin_brown_2010}.
An alternative approach is considered by \cite{lykou_ntzoufras_2013} who tune the Bayesian lasso penalty based on Pearson correlations at the limit of significance determined by Bayes factors.
Furthermore, fully Bayesian shrinkage methods include the horseshoe prior 
\citep{carvalho_etal_2010} and the double generalized Pareto \citep{armagan_etal_2012}.

In our case, the tuning parameter $\bl$ is vector valued and for the applications we consider we would like fast and efficient approaches by which to set it. 
To this end we propose three EB approaches for tuning $\bl$. Here, by EB we refer generically  to any procedure that uses the data to estimate  hyper parameters. 
We consider three specific estimators based on (i) minimizing the leave-one-out CV error, (ii) maximizing the marginal likelihood, and (iii) locating the maximum-a-posteriori (MAP) value under a data-dependent prior. 
All three are free of user input and are computationally fast.
We discuss each in turn.

\bigskip

\noindent \textbf{Leave-one-out cross-validation (CV) estimator:} 
The leave-one-out CV 
error in our case can be computed as
\begin{equation}
\widehat{\bl}_{\CV}  =\argmin_{\bl}\by^T(\I+\Gl)^{-1}\big[\diag(\I+\Gl)^{-1}\big]^{-2}(\I+\Gl)^{-1}\by .
\label{CV}
\end{equation}
This is similar to the well-known case of ordinary least squares; for proof see supplement, Appendix \ref{AppB}. 

\bigskip

\noindent \textbf{Marginal likelihood (ML) estimator:} 
A common EB approach is to use the marginal likelihood; in our case from Eq. \eqref{marg_lik} the  quantity to be maximized is
\begin{equation}
\widehat{\bl}_{\ML}  =\argmax_{\bl}|\I+\Gl|^{-\frac{1}{2}}\big[\by^T(\I+\Gl)^{-1}\by\big]^{-\frac{n}{2}}. 
\label{ML} 
\end{equation}

\bigskip

\noindent \textbf{Maximum-a-posteriori (MAP) estimator:} We consider a product-exponential data-dependent prior for $\boldsymbol{\lambda}$ with prior mode at zero, prior mean equal to $\widehat{\bl}_{\CV}$ as given in \eqref{CV}, and prior variance $\widehat{\bl}_{\CV}^2$, i.e.,   $\pi(\boldsymbol{\lambda})\equiv\prod_{k=1}^{K}\mathrm{Exp}\big(\widehat{\lambda}^{-1}_{k\CV}\big)$.
The rationale is that a smaller individual estimated penalty $\widehat{\lambda}_{k \CV}$ corresponds to a stronger belief that the corresponding $\Xj$ matrix contains useful signal and therefore to a smaller prior variance (especially when $\widehat{\lambda}_{k \CV}<1$). On the other hand as $\widehat{\lambda}_{k \CV}$ increases we let the quadratic prior variance account for the chance that there is actually some useful signal in $\Xj$ which passes undetected by the leave-one-out CV approach.   
The resulting posterior mode estimate is 
\begin{equation}
\widehat{\bl}_{\PM}  =\argmax_{\bl}|\I+\Gl|^{-\frac{1}{2}}\big[\by^T(\I+\Gl)^{-1}\by\big]^{-\frac{n}{2}}\exp\Bigg( -\sum_{k=1}^{K}\frac{\lambda_k}{\widehat{\lambda}_{k\CV}} \Bigg).
\label{PM}
\end{equation}

The optimization problems in Eqs. \eqref{CV}, \eqref{ML}, and \eqref{PM} are typically  well behaved, with the shape of the optimization landscape depending on relative signal strength.
For non-informative sources, it is relevant to note that the magnitude of $\lambda_k$ does not affect the ridge solution after certain large values (as all coefficients are shrunk to zero); 
hence, it is reasonable to simply define a large upper bound for the optimization.
A key point is that the number of available data sources $K$ will typically not be large, hence the vector $\bl$ is low dimensional, allowing the optimizations to be efficiently solved  via standard routines. 

\section{Sparse SBR }
\label{Sec:SSBR} 

The SBR posterior mode in \eqref{Lambda_mode} is non-sparse (``dense") in the sense that the regression coefficients will not be set to exactly zero. In this Section we propose a methodology for ``sparsifying" SBR. 
The idea is to find a sparse approximation to the full (dense) Bayesian solution that is closest to it in a Kullback-Leibler (KL) sense.
To do so, we minimize the KL divergence with respect to the posterior distribution of the regression vector, but subject to a lasso-type $\ell_1$ constraint to ensure sparsity. We show first a general solution that is suitable for small to moderate $p$ and then go on to consider a relaxed solution that is applicable to the large $p$ case. 
The solutions presented below bear a resemblance to other Bayesian post-processing approaches \citep{bondell_reich2012, hahn_carvalho2015} and to frequentist methods in the context of wavelet regression \citep{Antoniadis_Fan2001} and   ridge logistic regression for text categorization \citep{Aseervatham2011}. 
However, these are rooted in different arguments and  not equivalent to the KL-based approach below. The particular connection with the pCR solution of \cite{bondell_reich2012} is discussed in Section \ref{Sec: pCR}.

\subsection{Sparsification using the KL divergence}
\label{Sec: Gen_sol}

Let $f(\bb)\equiv\N_p(\bbhat,\Sb\sigma^2)$ denote the true posterior over $\bb$, conditional on $\sigma^2$, with mode and covariance as in Eqs. \eqref{Lambda_mode} and \eqref{Cov}, respectively, and let $q(\bb)\equiv \N_p(\g,\Sb\sigma^2)$ denote an approximate conditional posterior where $\g$ is the approximate mode (this will provide a sparsification of $\bbhat$). The idea is to minimize the KL divergence from $q$ to $f$ under an $\ell_1$ penalty on vector $\g$ to induce sparsity. It is easy to show that the KL divergence from $q(\bb)$ to $f(\bb)$ is
\begin{equation}
D_{\mathrm{KL}}(f||q)=\frac{1}{2\sigma^2}(\bbhat-\g)^T\Sb^{-1}(\bbhat-\g).
\label{kl}
\end{equation}   
Note that $D_{\mathrm{KL}}$  in \eqref{kl} is a true distance metric (satisfying non-negativity, symmetry and the triangle inequality).
Note also that the presence of the nuisance parameter $\sigma^2$ cannot be ignored when the minimization also involves a $\ell_1$ penalty on $\g$. In principle, one could work with the marginal posterior distribution of $\bb$ (a multivariate $t$ distribution) in order to avoid consideration of $\sigma^2$. However, in this case working with the KL divergence is not straightforward. Another option would be  to use a plug-in posterior point estimate in \eqref{kl}  such as the mode or mean of $\sigma^2$. Instead, here we pursue a tuning-free approach in which $\sigma^2$ is integrated out; specifically, we work with
\begin{equation}
\mathbb{E}_{\sigma^2|\by}\left[D_{\mathrm{KL}}(f||q)\right]=\int D_{\mathrm{KL}}(f||q)p(\sigma^2|\by) \, \mathrm{d}\sigma^2 =
\frac{c_{n,\bl}}{2}(\bbhat-\g)^T\Sb^{-1}(\bbhat-\g),
\label{ekl}
\end{equation}
with the posterior of $\sigma^2$ given in \eqref{post_sigma}
and 
$c_{n,\bl}=n \, q_{\bl}^{-1}$,
where $q_{\bl}=\by^T(\I+\Gl)^{-1}\by$. 
Using the posterior mean or mode of $\sigma^2$ results in $c_{n,\bl}=(n-2)q_{\bl}^{-1}$ and $c_{n,\bl}=(n+2)q_{\bl}^{-1}$, respectively.
Using \eqref{ekl}, the general solution including the $\ell_1$ penalty is
\begin{equation}
\ghat=\argmin_{\g}\frac{c_{n,\bl}}{2}(\bbhat-\g)^T\Sb^{-1}(\bbhat-\g) + \alpha\lVert\g\rVert_1,
\label{G_Sol}
\end{equation}      
where $\alpha>0$ controls the sparsity of $\widehat{\g}$. Clearly, the SSBR solution implies a lasso-type model with the particularity of a saturated design where the analogue to sample size equals $p$. 
This means that SSBR can include at most $p$ predictors (unlike classical lasso which for $p>n$ can include at most $n$ predictors).
The minimization in \eqref{G_Sol} can be solved as a lasso problem by setting $\by^*=\sqrt{c_{n,\bl}}\Sb^{-1/2}\bbhat$ as response variable and $\X^*=\sqrt{c_{n,\bl}}\Sb^{-1/2}$ as design matrix, and using the efficient implementation of \verb|glmnet| in R \citep{R}.
Note, however, that this involves first calculating and then performing computations with the inverse covariance matrix, which can be problematic even for moderately large $p$.

As an alternative, we present a mathematically equivalent representation of \eqref{G_Sol} that is easier to work with. 
First, observe that the posterior covariance  in \eqref{Cov} can be written as $\Sb=\La^{-1/2}(\Ip-\M^T\M)\La^{-1/2}$, where $\M=(\I+\Gl)^{-1/2}\X\La^{-1/2}$. Now, consider the singular value decomposition (SVD) $\M=\U\D\V^T$, where $\U\in\mathbb{R}^{n\times n}$, $\D\in\mathbb{R}^{n\times p}$ and $\V=[\V_1 ~ \V_2]$ with $\V_1\in\mathbb{R}^{p\times n}$ and $\V_2\in\mathbb{R}^{p\times (p-n)}$. Note that computation of $\U,\D,\V_1$ is $\mathcal{O}(n^2p)$  and also that the last $p-n$ columns of $\D$ contain only zeros. In addition, the decomposition of $\M$ can be greatly sped up by first computing the SVD of $\M\M^T$. Some algebra (details provided in Appendix \ref{AppC} of supplementary material) reveals that the resulting equivalent SSBR solution requires as additional input only the $n$ singular values of $\M$ and matrix $\V_1$; specifically, the solution is    
\begin{align}
\ghat  = \argmin_{\g}  \frac{c_{n,\bl}}{2}\lVert \tilde{\D}^{1/2}\V_1^T\La^{1/2}(\bbhat-\g)\rVert_2^2+  
\frac{c_{n,\bl}}{2}\lVert\La^{1/2}(\bbhat-\g)\rVert_2^2 + \alpha\lVert\g\rVert_1,
\label{G_FastSol}
\end{align}
where $\tilde{\D}$ is $n\times n$ diagonal with elements $\tilde{d}_i=d_i^2/(1-d_i^2)$ and $d_i$ are the singular values.
In practice \eqref{G_FastSol} can be solved as an augmented-lasso problem, using 
\begin{equation*}
\by^*=\sqrt{c_{n,\bl}}\begin{pmatrix}
{\tilde{\D}^{1/2}\V_1^T\La^{1/2}\bbhat} \\
\La^{1/2}\bbhat
\end{pmatrix} \quad \mbox{and} \quad
\X^*=
\sqrt{c_{n,\bl}}\begin{bmatrix}
\tilde{\D}^{1/2}\V_1^T\La^{1/2} \\
\La^{1/2}
\end{bmatrix}
\end{equation*}        
which have respective dimensionality $(n+p)\times 1$ and $(n+p)\times p$. The matrix multiplication involves mainly sparse matrices, as $\tilde{\D}$ and $\La$ are diagonal. An interesting side note is that such SVD decompositions can be generally used in $p>n$ problems for covariances of the form $(\X^T\X+\mathbf{Z})^{-1}$ for some diagonal matrix $\mathbf{Z}$. 
Thus, using \eqref{G_FastSol} instead of \eqref{G_Sol} is advocated when $p$ is large. However, for applications with ultra-large $p$ the approach presented in Section \ref{Sec: Rel_sol}, which leads to a closed-form expression for $\ghat$, is the only practical option.  


\subsection{Relation to penalized credible regions}
\label{Sec: pCR}

The pCR approach \citep{bondell_reich2012} seeks solutions of the form $\bbhat=\argmin_{\bb}\Vert\bb\Vert_0$ subject to $\bb \! \in \! \mathcal{C}_\alpha$, where $\mathcal{C}_\alpha$ is the $(1-\alpha)\times100\%$ credible set. Under a normal ridge-prior on $\bb$ and a inverse-gamma prior on $\sigma^2$ this translates to a feasible set  of the form $\{\bb:(\bb-\bbhat)^T\Sb^{-1}(\bb-\bbhat)\le C\}$ for some $C$ corresponding to a specific credible region.

To tackle the obvious computational challenges of the above solution, the authors initially relax the $\ell_0$ norm to a smooth homotopy between $\ell_0$ and $\ell_1$, and subsequently apply a local linear approximation which results in a convex $\ell_1$ optimization problem. The resulting solution is very similar to our solution obtained through the KL approach; in fact by setting the penalty in \eqref{G_Sol} as $\alpha=\frac{c_{n,\bl}}{2}\lVert \bbhat\rVert_1^{-2}\xi$ (now $c_{n,\bl}/2$ no longer affects the optimization and $\xi$ is the new penalty) we recover exactly the pCR solution. As noted in \cite{bondell_reich2012} there is a one-to-one correspondence between $C$ and $\xi$; however, it is highly non linear. Under this setting we have selection consistency (under mild regularity conditions) when $p$ is fixed or as long as $p/n\rightarrow 0$ for $n\rightarrow \infty$. The authors also demonstrate selection consistency (under stricter conditions) for rates $\log p=\mathcal{O}(n^c)$ for some $c\in(0,1)$ using univariate thresholding rules on simple ridge estimates.

In practice, the $\xi$ that corresponds to the asymptotically consistent sparse set is not recoverable. Theoretically, the penalty should depend on sample size so that $\xi_n\rightarrow 0$ faster than the posterior distribution concentrates around the ``true'' parameter value; however, under finite samples the selection of $\xi$ crucially affects the sparsity of $\bb$ \citep{hahn_carvalho2015}. Therefore, tuning of $\xi$ is handled through  common grid search and inspection of regularization plots/prediction errors in \cite{bondell_reich2012}.


\subsection{A relaxed solution for the very large-$p$ case}
\label{Sec: Rel_sol}
Instead of the KL divergence to the posterior used above, consider the  KL divergence between the quantities 
$q^*(\bb) =   \N_p(\g,\Vb\sigma^2)$ and $f^*(\bb)  =   \N_p(\bbhat,\Vb\sigma^2)$, 
with $\Vb=\diag(\Sb)$.  This amounts to setting as target distribution the product of the marginal posterior densities. 
The use of independent posterior factorizations is common in various settings; for instance, in marginal likelihood estimation \citep{Botev_2013,perrakis_marglik}, in expectation-propagation algorithms \citep{Minka_2001}, and in variational Bayes  \citep{Bishop}.

Working with the diagonal matrix $\Vb$ leads to the following  minimization 
\begin{align}
\ghat= & \argmin_{\g}\frac{c_{n,\bl}}{2}(\bbhat-\g)^T\Vb^{-1}(\bbhat-\g) + \alpha\lVert\g\rVert_1 \nonumber \\ 
= & \argmin_{\g} \sum_{j=1}^{p}\frac{c_{n,\bl}}{2}(\widehat{\beta}_j-\gamma_j)^2v_{j}^{-1}+\alpha|\gamma_j|,
\label{R_Sol}
\end{align}
where $v_{j}$ is the $j$-th element, for $j=1,\dots,p$, of the main diagonal of $\Vb$. Note that the main diagonal elements are feasible to calculate even for very large $p$; this can be achieved by calculating  Eq. \eqref{parallel} in parallel. Moreover, the minimization in \eqref{R_Sol} has a closed-form solution which is as follows
\begin{equation}
\widehat{\gamma}_j=\begin{cases}
\displaystyle
\widehat{\beta}_j-\mathrm{sign}(\widehat{\beta}_j)\frac{q_{\bl}}{n}v_{j}\alpha &, ~ \mbox{if} ~|\widehat{\beta}_j|> \frac{q_{\bl}}{n}v_{j}\alpha\\
0 &, ~ \mbox{otherwise}.
\end{cases}
\label{relaxed}       
\end{equation} 
The derivation of \eqref{relaxed} is provided as supplementary material (Appendix \ref{AppD}).

Note that for fixed $p$ and $n\rightarrow\infty$ we obtain $\widehat{\gamma}_j=\widehat{\beta}_j$
which makes sense from an asymptotic perspective. 
However, when $\alpha$ is a constant not depending on $n$ either directly or indirectly (e.g., through $\widehat{\beta}_j$),  a ``non-sparsifying'' effect may be triggered even for moderate sample size, which is in contrast to our initial intent. Setting of $\alpha$ is discussed next. 

\subsection{Tuning of $\alpha$}


Specification of $\alpha$ can be handled via a grid search with the aim to find the $\alpha$ that minimizes a specific criterion. This is expected to be relatively fast using the general solution for small/moderate $p$, while under the relaxed solution, once the variances are calculated, the grid search requires only checking a true/false statement. This strategy will typically produce a full path of solutions which can also be used to produce regularization plots. We acknowledge this as a valid common strategy, however, we do not further pursue it here. Instead, we consider a faster, tuning-free, alternative which borrows information from the SBR solutions.

We consider parameter-specific and source-specific adaptive penalties for each $\widehat{\gamma}_{jk}$, where $j=1,\dots,p_k$ and $k=1,\dots,K$. Specifically, we consider penalties of the form 
\begin{equation}
\alpha_{jk}=\Bigg(\frac{1}{|\widehat{\beta}_{jk}|}\Bigg)^{w_k},
\label{kappa}
\end{equation}
similar to the adaptive lasso approach \citep{Zou2006}.
The rationale in \eqref{kappa} is that the larger the magnitude of $\widehat{\beta}_{jk}$, the smaller the corresponding penalty. 
In addition, we restrict to $w_{j}\in(0,1)$ which leads to reasonable shrinkage when $|\widehat{\beta}_{jk}|>1$ and avoids extreme shrinkage when $|\widehat{\beta}_{jk}|<1$.
Here, we consider one specific possibility, namely to treat the $w_k$'s as power-weights,  setting them equal to
\begin{equation}
w_k=\frac{\widehat{\lambda}_k}{\sum_{l=1}^{K}\widehat{\lambda}_l}.
\label{weights}
\end{equation}
The power weight quantifies the ``importance'' of a data source in relation to the others. Values close to zero and one indicate sources of ``high'' and ``low'' importance, respectively. Coefficients with an absolute value smaller than one (the common case) are penalized more in low-importance sources, while shrinkage on large coefficients (greater than one in absolute value) is relatively mild, and at most approximately equal to one (when $w_k\rightarrow 0$).
With this approach we take advantage of the available information from the previous SBR step, i.e., parameter-specific shrinkage through $\widehat{\beta}_{jk}$ and source-specific shrinkage through $\widehat{\lambda}_k$. Arguably, this strategy may result in undesirable non-sparse solutions, but that will be in the rare, and rather unrealistic, case where $K$ is large and all sources are equally important in the sense that the $\widehat{\lambda}_k$'s will be more or less the same; a setting where in fact a single $\lambda$ SBR approach is more suitable.

Note that, to implement this approach under the general solution in \eqref{G_Sol} we find first $\boldsymbol{\gamma}^*$ via \verb|glmnet| (with penalty set to one) using as design matrix $\X^*=\sqrt{c_{n,\bl}}\Sb^{-1/2}\mathbf{A}$, where $\mathbf{A}$ is the diagonal matrix with elements the reciprocals of \eqref{kappa}. The solution is then $\ghat=\mathbf{A}\boldsymbol{\gamma}^*$.

\begin{table}
	\centering
	\scalebox{.79}{\begin{tabular}{l l l l}
			\hline 
			Method & Characteristics & Issues & Proposed Solutions \\
			\hline 
			\multirow{3}{*}{SBR}& Scalable to very large $p$&  \multirow{3}{*}{Tuning of $\bl$}& CV estimator, Eq. \eqref{CV}\\
			&  Very fast &  & ML estimator, Eq. \eqref{ML}\\
			&  Dense solutions &  & MAP estimator, Eq. \eqref{PM}\\
			\hline 
			\multirow{3}{*}{SSBR} & Scalable to moderately large $p$ & \multirow{3}{*}{
				Tuning of $\alpha$
			}  & \multirow{3}{*}{
				Tune via Eqs. \eqref{kappa} \& \eqref{weights}} \\
			& Relatively fast &  & \\
			& Sparse solutions &  & \\
			\hline
			\multirow{3}{*}{Relaxed SSBR} & Scalable to very large $p$ & \multirow{3}{*}{
				\begin{minipage}[l]{0.27\textwidth}
					\begin{description}
						\setlength\itemsep{0.01em}
						\item[1)] Tuning of $\alpha$
						\item[2)] Control effect of $n$
					\end{description}
				\end{minipage}
			} & \multirow{3}{*}{
				\begin{minipage}[c]{0.35\textwidth}
					\begin{description}
						\setlength\itemsep{-0.5em}
						\item[1)] Tune via Eqs. \eqref{kappa} \& \eqref{weights} 
						\item[2)]  $f_n=1$ (no control) or  
						\item[]    ~~ $f_n=\log n$ (cSSBR)
					\end{description}
				\end{minipage}
			}\\
			& Fast &  & \\
			& Sparse solutions &  & \\
			\hline 
	\end{tabular}}
	\caption{Overview of methods, characteristics, issues and solutions.}
	\label{newmethods}
\end{table}

As a final comment, we remark that despite the fact that this penalization approach depends indirectly on sample size through the regression coefficients in \eqref{kappa} and the shrinkage parameter in \eqref{weights}, it may still be sensitive to the ``non-sparsifying'' effect on the relaxed SSBR solution discussed at the end of Section \ref{Sec: Rel_sol}. Controlling this effect requires scaling the penalty in \eqref{kappa} by a factor $f_n=f(n)$; however, automatic tuning of $f_n$ is not straightforward. Empirical results (see Section \ref{SIM_STUDY}) suggest that $f_n=\log(n)$ can lead to a reasonable balance between sparsity and predictive performance. We will call this 
relaxed extension  ``controlled" SSBR or cSSBR (as it ``controls" for sample size). 
Table \ref{newmethods} provides an overview of the different methods under consideration, their characteristics, the issues that arise under each approach, and our proposed solutions.

\section{Simulation study}
\label{SIM_STUDY}

In this Section we present a simulation study aimed at 
mimicking  a typical modern biomedical application involving multiple data types.
Reflecting the relative ease with which multiple data modalities can now be acquired such designs are becoming common, with examples including the Cancer Genome Atlas (\url{https://cancergenome.nih.gov}), the Alzheimer's Disease Neuroimaging Initiative (\url{http://adni.loni.usc.edu}), and the Rhineland Study (\url{http://www.rheinland-studie.de}), among many others.

\subsection{Set-up}
{\bf The problem}. We consider a regression problem with covariates from three sources, namely clinical (CL), gene-expression (RNA), and genetic (single nucleotide polymorphism or SNP) data with respective (simulated) feature matrices $\mathbf{X}_{\CL}$, $\mathbf{X}_{\RNA}$ and $\mathbf{X}_{\SNP}$.
The number of covariates in each data source is set equal to $p_{\CL}=26$, $p_{\RNA}=2000$ and $p_{\SNP}=100000$. 
Although the methods we propose can cope with larger  $p$, we restrict total $p$ in this Section to facilitate empirical  comparison with standard methods.

\medskip
\noindent
{\bf Covariates}. The covariate matrices for the clinical and gene-expression variables are generated as $\mathbf{X}_{\CL}\sim\N_{p_{\CL}}(\mathbf{0},\boldsymbol{\Sigma}_{\CL})$ and
$\mathbf{X}_{\RNA}\sim\N_{p_{\RNA}}(\mathbf{0},\boldsymbol{\Sigma}_{\RNA})$, respectively.   
Here $\boldsymbol{\Sigma}_{\CL}$ and $\boldsymbol{\Sigma}_{\RNA}$ are covariance matrices estimated from (real) phenotype and gene-expression data from the \textit{Drosophila} Genetic Reference Panel (DGRP) \citep{drosophila} (data available online at \url{http://dgrp2.gnets.ncsu.edu/data.html}).    
To simulate the genetic data  $\mathbf{X}_{\SNP}$ we use a  block-diagonal covariance structure.
We specify $\boldsymbol{\Sigma}_{\SNP}=\diag(\boldsymbol{\Sigma}_{\SNP}^{1},\dots,\boldsymbol{\Sigma}_{\SNP}^{B})$, where each $\boldsymbol{\Sigma}_{\SNP}^{b}$ is of size $S \times S$ (with $S=p_{\SNP}$/B) and is generated from a inverse-Wishart  with $S$ degrees of freedom and identity scale matrix, i.e.,  $\boldsymbol{\Sigma}_{\SNP}^{b}\sim \mathrm{IW(S,\boldsymbol{I}_{S})}$ for $b=1,\dots, B$.
As $\mathbf{X}_{\SNP}$ dominates in terms of dimensionality the specification of $B$ essentially controls the overall correlation level. We consider two simulation scenarios: (i) $B=1000$ corresponding to 1000 blocks of size 100 (``low-correlation scenario") and (ii) $B=100$ corresponding to 100 blocks of size 1000 (``high-correlation scenario").     
We first generate $\mathbf{X}^c_{\SNP}\sim\N_{p_{\SNP}}(\mathbf{0},\boldsymbol{\Sigma}_{\SNP})$
and then discretize 
in correspondence to the common SNP encoding 0/1/2 (homozygous major allele/heterozygous/homozygous minor allele). The discretization is tuned in order to give a reasonable empirical distribution of SNPs; specifically, 
for $j=1,\dots,p_{\SNP}$
we discretize as
\begin{equation*}
X_{j \SNP}=\begin{cases}
0, &~ \mbox{if} ~ |X^c_{j \SNP}|<1.5, \\
1, &~ \mbox{if} ~ |X^c_{j \SNP}|\ge 1.5 ~ \mbox{and} ~  |X^c_{j \SNP}|<2.5,\\
2, &~ \mbox{if} ~ |X^c_{j \SNP}|\ge 2.5.
\end{cases}
\end{equation*}

\medskip
\noindent
{\bf Regression coefficients and sparsity}. 
For the (true) regression vectors $\bb_{\CL}$, $\bb_{\RNA}$ and $\bb_{\SNP}$ we consider the following levels of sparsity (fraction of non-zero $\beta$'s); $s_{\CL}=50\%$, $s_{\RNA}=5\%$ and $s_{\SNP} \in \{1\%,10\%,50\%\}$. Varying  sparsity of $\bb_{\SNP}$ gives three scenarios for overall sparsity $s$: (i) $s\approx 1\%$ (sparse scenario), (ii) $s\approx 10\%$ (medium scenario) and (iii) $s\approx 50\%$ (dense scenario).
Let $p^*_{\CL}$, $p^*_{\RNA}$ and $p^*_{\SNP}$ denote the respective number of elements in the sub-vectors $\bb^*_{\CL}$, $\bb^*_{\RNA}$ and $\bb^*_{\SNP}$ containing the non-zero beta coefficients. 
The non-zero betas are generated from the generalized normal distribution (GND). Following the parameterization in \citet{Mineo_2003} the probability distribution function of a GND($\mu,\sigma,u$) with location $\mu\in \mathbb{R}$, scale $\sigma>0$, and shape $u>0$ is given by 
\begin{equation*}
f(x)=\frac{1}{2u^{1/u}\sigma\Gamma(1+1/u)}\exp \Bigg(-\frac{|x-\mu|^u}{u\sigma^u}\Bigg).
\end{equation*}
The GND includes as special cases the normal ($u\! = \!2$) and the double exponential ($u\! = \!1$) distributions. To avoid these particular cases (which could potentially bias the simulation towards ridge or lasso respectively) we set $u \! = \!1.5$ and generate the non-zero effects as $\beta^*_{j \CL}\sim\mathrm{GND}(0,\widehat{\sigma},1.5)$, for $j=1,\dots,p^*_{\CL}$,
$\beta^*_{j \RNA}\sim\mathrm{GND}(0,\widehat{\sigma},1.5)$, for $j=1,\dots,p^*_{\RNA}$, and
$\beta^*_{j \SNP}\sim\mathrm{GND}(0,2\widehat{\sigma}/3,1.5)$, for $j=1,\dots,p^*_{\SNP}$.
The signal strength is controlled via the scale parameter $\widehat{\sigma}$ (this is downscaled by a factor of 1.5 for the SNP coefficients to control the total amount of signal in the SNPs). 
To complete the specification of the simulation we set this scale parameter 
by considering the finite-sample risk in a simplified CL-only oracle-like setup. Specifically we consider 
the  correlation induced between predictions $\mathbf{X}^{\mathrm{test}}_{\CL}\bbhat_{\CL}$ (under the OLS estimate using the low-dimensional CL data only) and out-of-sample test data (with the data-generating mechanism being a linear model with conditional mean $\mathbf{X}_{\CL}\bb_{\CL}$ and error variance equal to unity). 
Specifically, we set $\widehat{\sigma}=0.1$ which results in an average out-of-sample correlation of 0.6 when $n$=100 and $n_{\mathrm{test}}$=5000.  

Given the above configurations (low/high correlation and sparse/medium/dense scenarios) we generate data from the model
\begin{equation*}
y_i=\mathbf{x}_{i \CL}^T\bb_{\CL}+\mathbf{x}_{i \RNA}^T\bb_{\RNA}+\mathbf{x}_{i \SNP}^T\bb_{\SNP}+\varepsilon_i,
\end{equation*}
where $\varepsilon_i\sim N(0,1)$ and $i=1,\dots,n_{\mathrm{train}}$ with $n_{\mathrm{train}}\in\{100,250,500\}$. The test sample size $n_{\mathrm{test}}$ always equals 5000. Each simulation scenario is repeated 50 times.

\medskip
\noindent
{\bf  Methods under comparison}.
We consider SBR and the corresponding sparse extensions (unless otherwise noted, SSBR/cSSBR will refer to the relaxed approach discussed in Section \ref{Sec: Rel_sol}).
Specifically, we consider SSBR with the penalty terms in \eqref{kappa} and the power-weights in \eqref{weights}, as well as cSSBR approaches with $f_n^{(1)} {=} \sqrt n$, $f_n^{(2)} {=} \log(n)$ and $f_n^{(3)}  {=} \sqrt{\log(n)}$. 
Results reveal no significant differences with respect to the EB $\bl$ estimates (proposed in Section \ref{tuning}) under SBR. However, the MAP estimator resulted in better performance under the sparse approaches and, therefore, here we focus on this approach. Corresponding results under the CV and ML approaches are provided in Appendix \ref{AppE}.
Furthermore, we present results obtained from $f_n^{(2)}$ as this option led to a good balance between sparsity and predictive performance.  
We compare to standard ridge, elastic net (enet), and lasso (with $\lambda$ set to minimize the mean squared error from 5-fold CV using package \verb|glmnet| in R). The enet control parameter $\alpha$ was tuned via grid search over the interval $[0.1,0.9]$ using a step of 0.1.

\subsection{Results}

\begin{figure}[h]
	\centering{}\includegraphics[width=\columnwidth]{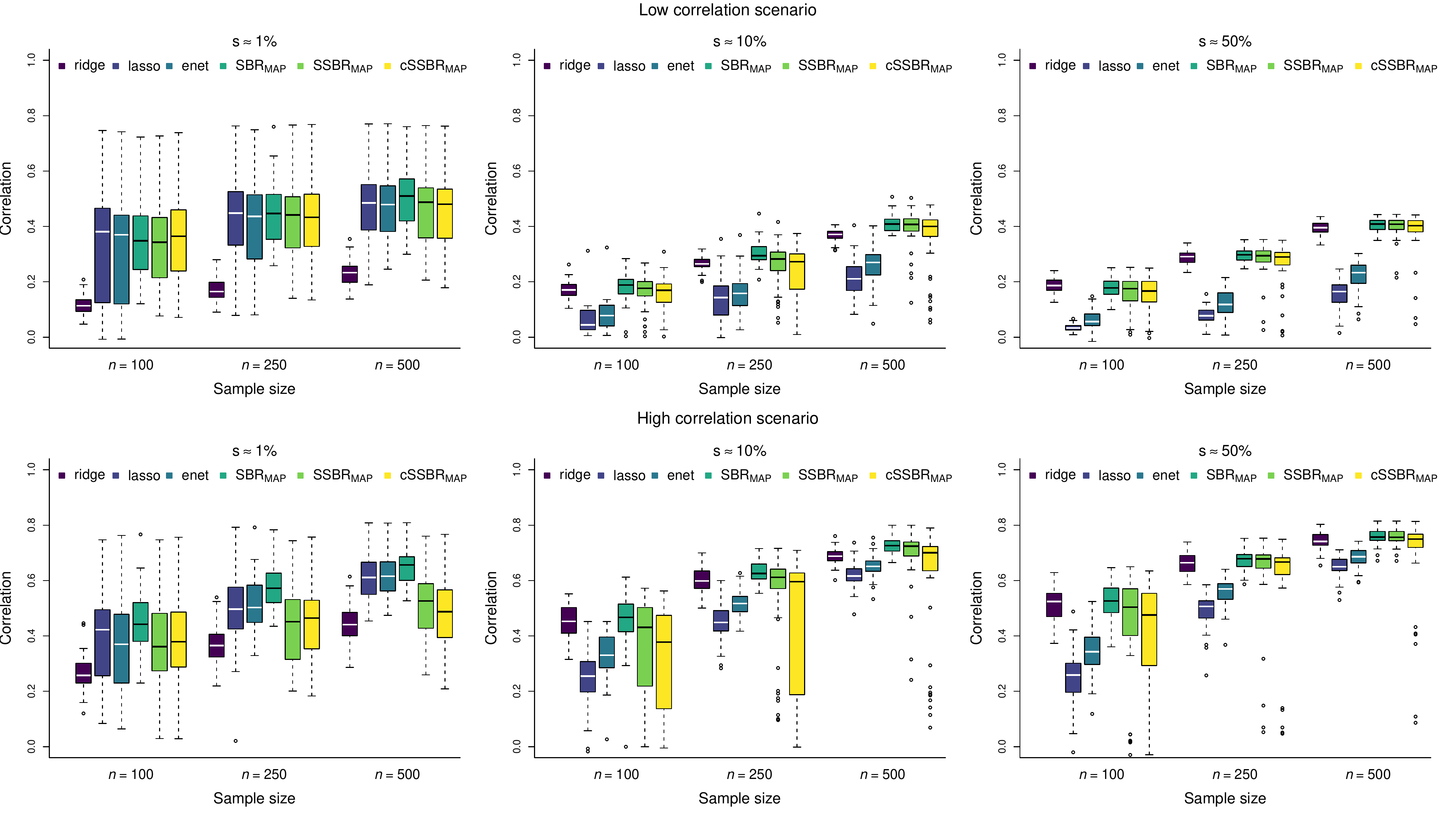}
	\caption{Simulation study. Comparison of ridge, lasso and elastic net to SBR, SSBR and cSSBR methods under the MAP estimator for the low correlation (top panels) and high correlation (bottom panels) scenarios. Each panel shows correlations between predictions and held-out test data at various training sample sizes and sparsity levels as indicated. Boxplots are over 50 sampled datasets. The total dimension $p$ is over 100,000 with three (simulated) data sources (see text for details).}
	\label{cor_low}
\end{figure}

Boxplots of out-of-sample correlations between predictions and test data under the low-correlation and high-correlation simulations are presented in Figure \ref{cor_low}.
Results for (classical) ridge and lasso are as expected; lasso clearly performs better in the sparse case, while ridge does better in the medium and dense cases. Also, more or less as expected enet performs as well as lasso in the sparse scenario and slightly better than lasso in the medium and dense scenarios; although in the latter two cases it is outperformed by  ridge.  

SBR performs generally well. Specifically, we see that:
\begin{itemize}
	\setlength\itemsep{0.01em}
	\item In the sparse scenario SBR is nearly equivalent to lasso/enet. 
	\item In the medium scenario SBR generally outperforms ridge, lasso, and enet.
	\item In the dense scenario SBR is nearly equivalent to ridge.
\end{itemize}

For SSBR approaches we see:
\begin{itemize}
	\setlength\itemsep{0.01em}
	\item Under low correlations SSBR/cSSBR are slightly worse than SBR but competitive.
	\item Under high correlations in the medium/dense cases with large $n$ SSBR/cSSBR are competitive to SBR. 
\end{itemize}

\begin{figure}
	\centering{}\includegraphics[width=\columnwidth]{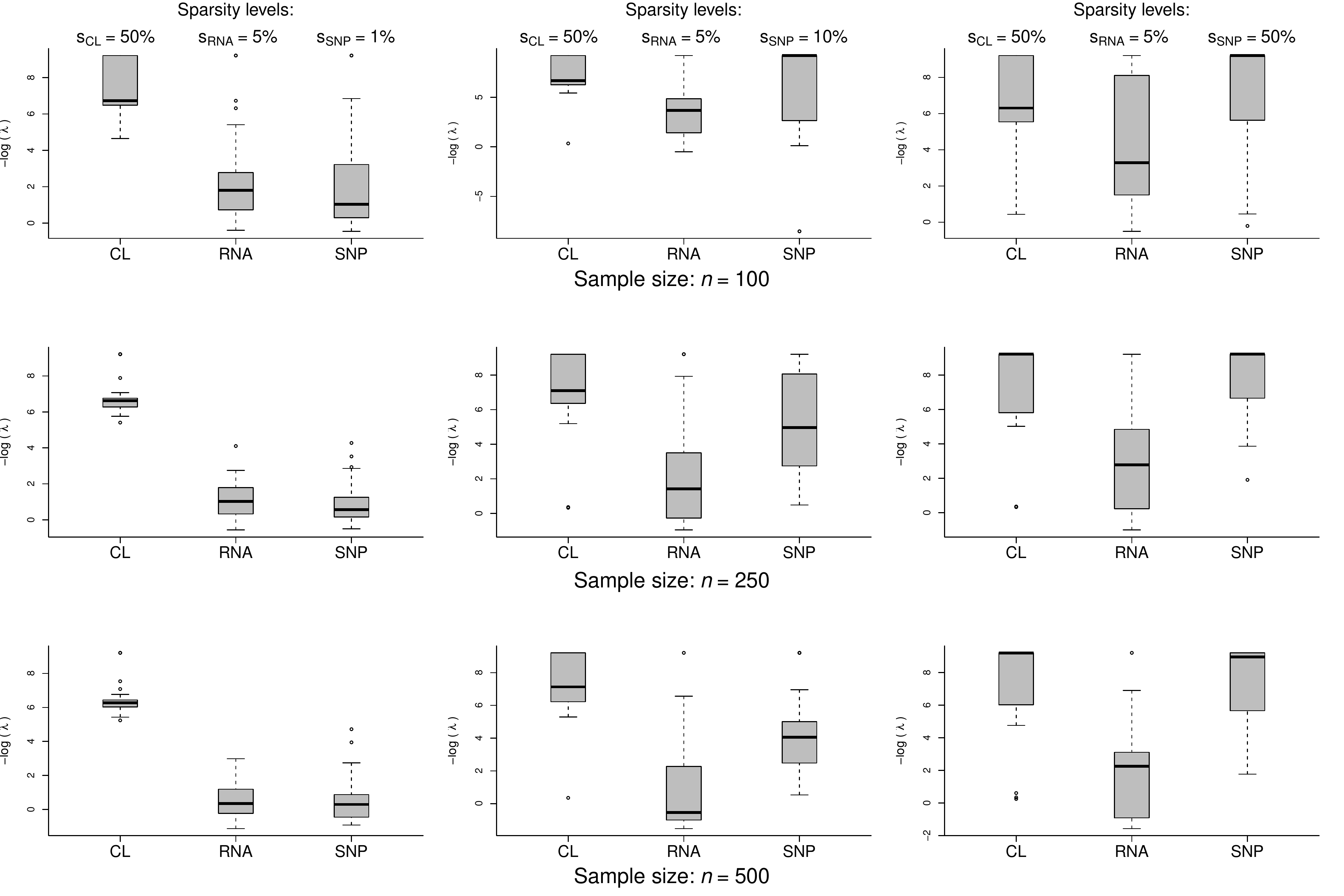}
	\caption{Simulation study, source-specific shrinkage levels. Boxplots showing source-specific $-\log\hat\lambda$ values from SBR based on the MAP estimator in the low-correlation simulations under various levels of sparsity for $n=100$ (top row), $n=250$ (middle row) and $n=500$ (bottom row).}
	\label{lambdas_low}
\end{figure}

Figure \ref{lambdas_low} shows the resulting values of $-\log\widehat{\lambda}_k$ for $k=\{\mathrm{CL},\mathrm{RNA},\mathrm{SNP}\}$ (higher values correspond to lower penalty; i.e., higher estimated importance) in the low-correlation simulations, and provides useful insights concerning the behaviour of SBR methods.
The estimates appear to adjust well with appropriate source-specific penalization. This adaptation  allows SBR to perform well when dealing with multiple data sources.
The corresponding plots from the high-correlation scenarios (not shown) are very similar.

As noted above, the SSBR (without or with control) solutions seem to allow equally good predictive performance, in certain cases, as the dense SBR solutions.
In addition, they employ fewer parameters; Table \ref{spars} shows the average sparsity (over the 50 repetitions) induced by SSBR methods under the various simulations. 
As seen, the solutions appear to adjust to the true underlying sparsity. In addition, controlling for the effect of sample size yields much sparser models.
In contrast, lasso and elastic net 
produced very sparse models that showed no such adaptation; see Appendix \ref{AppE} of supplementary material for detailed results. We note that lasso can include at most $n$ predictors and here yielded extremely sparse solutions (ranging from 0.008\% to 0.45\%). Elastic net can in principle include more predictors than observations, but in our simulations it did not adjust to the underlying sparsity levels (ranging from 0.06\% to 1.2\%).

\begin{table}
	\centering{\scalebox{0.9}{\begin{tabular}{l|c|cccccc}
				\hline
				\multirow{3}{*}{Method} & \multirow{3}{*}{Sample size} & \multicolumn{6}{c}{Simulation scenario}  \\
				&  & \multicolumn{3}{c}{Low-correlation} & \multicolumn{3}{c}{High-correlation} \\
				&                         &     $s\approx 1\%$       &      $s\approx 10\%$        &     $s\approx 50\%$     &     $s\approx 1\%$       &      $s\approx 10\%$        &     $s\approx 50\%$\\ \hline
				\multirow{3}{*}{SSBR$_\PM$}  & $n=100$ &    14\%        &     47\%       & 
				57\%                       &      8\%    &    42\%        &     53\% \\
				& $n=250$ &    10\%        &     50\%       &
				71\%                        &      4\%    &    57\%        &     73\% \\
				& $n=500$ &    14\%        &     69\%       &
				77\%                        &      13\%    &    70\%        &     79\% \\ \hline
				\multirow{3}{*}{cSSBR$_\PM$}  & $n=100$ &    4\%        &     18\%       & 
				25\%                       &      2\%    &    16\%        &     22\% \\
				& $n=250$ &    2\%        &     24\%       &
				35\%                        &      1\%    &    25\%        &     38\% \\
				& $n=500$ &    4\%        &     35\%       &
				43\%                        &      3\%    &    33\%        &     42\% \\ \hline
	\end{tabular}}}
	\caption{Simulation study, induced sparsity. Average sparsity induced by the SSBR and cSSBR methods based on the MAP estimator over 50 repetitions of the low and high correlation simulations under varying true sparsity ($s$).}
	\label{spars}
\end{table}

\begin{figure}[H]
	\centering{}\includegraphics[scale=0.45]{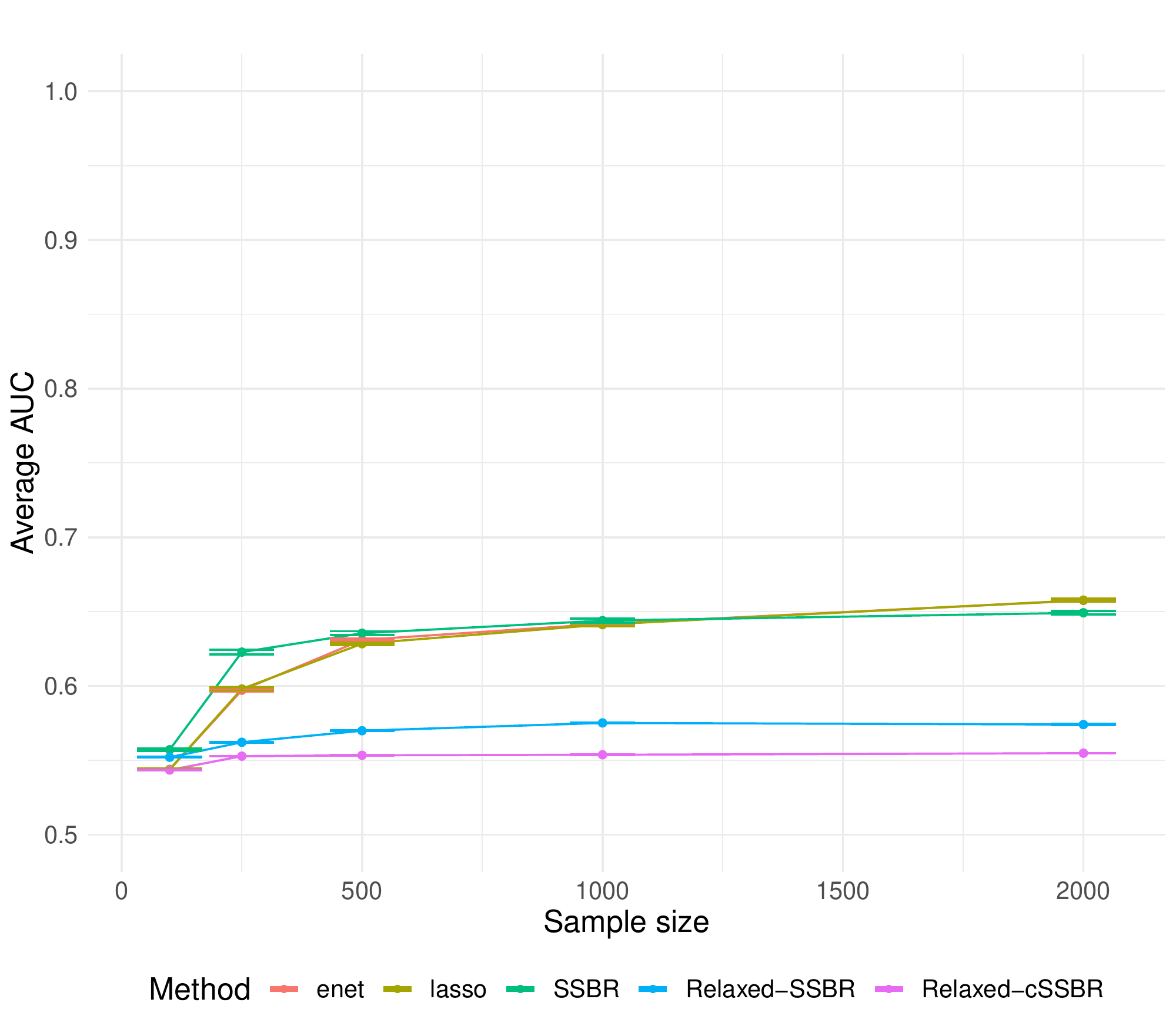}
	\caption{Simulation study, variable selection. One standard error plots of the average area under the ROC curve (AUC), over 20 repetitions, as a function of sample size.}
	\label{var_sel}
\end{figure}

The focus of this paper is on prediction, but it is interesting to consider the variable selection behavior of the proposed methods. As already noted, lasso and enet yielded extremely sparse solutions for the problem considered above
and are not suitable for selection in this particular $p \gg n$ setting.
We therefore considered a smaller problem which includes only the simulated CL and RNA data with $p=2026$ and $s\approx 6\%$. 
We summarize our findings via the area under the ROC curve (AUC; calculated using the absolute values of regression coefficients). 
Results comparing lasso, enet, SSBR and relaxed-SSBR/cSSBR (all under the MAP estimate) are shown in Figure \ref{var_sel}. Results based on the  CV and ML estimates (not shown) are similar. 
We see that SSBR in this case is competitive with the lasso and enet.



\subsection{Computational performance}
We conclude by examining computational burden as a function of total dimension $p$ and in comparison with the lasso. To do so we include a fourth ``data source'' (which is simply Gaussian noise)
that adds a variable number of predictors. The number of these additional predictors   is set so that the total $p\in\{5 {\times} 10^5,\, 10^6,\, 2.5 {\times}10^6, \, 5 {\times} 10^6,\, 10^7\}$. Sample size is set equal to 100. 
Computations were carried out on a compute server with 128 cores (2.28GHz) and 1TB of RAM.        
For lasso we treat the binding of individual matrices into one data matrix (an operation not needed for SBR) as a pre-processing step  and do not include this  in the reported runtimes. 
We consider two approaches for tuning the lasso penalty. The first is the most commonly used approach in practice; i.e., cross-validation (lasso$_{\CV}$). For this we use the parallel option in \verb|glmnet| for estimation of the penalty parameter via 10-fold CV (the default option).
The second approach is to do a grid search (lasso$_{\mathrm{grid}}$) with no CV. This is sometimes used in practice with the purpose of finding a value that maximizes a specific criterion (for instance BIC). 
We do not consider any particular criterion but report only the time needed to evaluate the lasso over the grid (i.e., we do not include any computational cost for the assessment  of any criterion).
For the sake of comparison we define a rough grid over the interval $[0.1,1)$ with a step of 0.2.
We note that in a practical application of the lasso the grid might need to be finer and the computational costs of assessing any statistical criterion might be nontrivial given the size of the matrices. These factors would increase the computational time needed for the lasso. 
For our methods we consider the SBR$_{\PM}$ approach, which requires evaluation of  both $\hat{\bl}_{\CV}$ in \eqref{CV} and $\hat{\bl}_{\PM}$ in \eqref{PM} (and is thus in principle the slowest), and also its corresponding relaxed sparse solution. We include the formation of transpose matrices and calculation of Gram matrices in reported runtime (although these could be regarded as pre-processing steps). We do not include ridge because it can be seen as a special case of SBR and hence would be equally fast when implemented as described here. 
We also do not include elastic net as this method will be slower than lasso due to tuning of the additional parameter $\alpha$.

\begin{figure}[]
	\centering{}\includegraphics[scale=0.4]{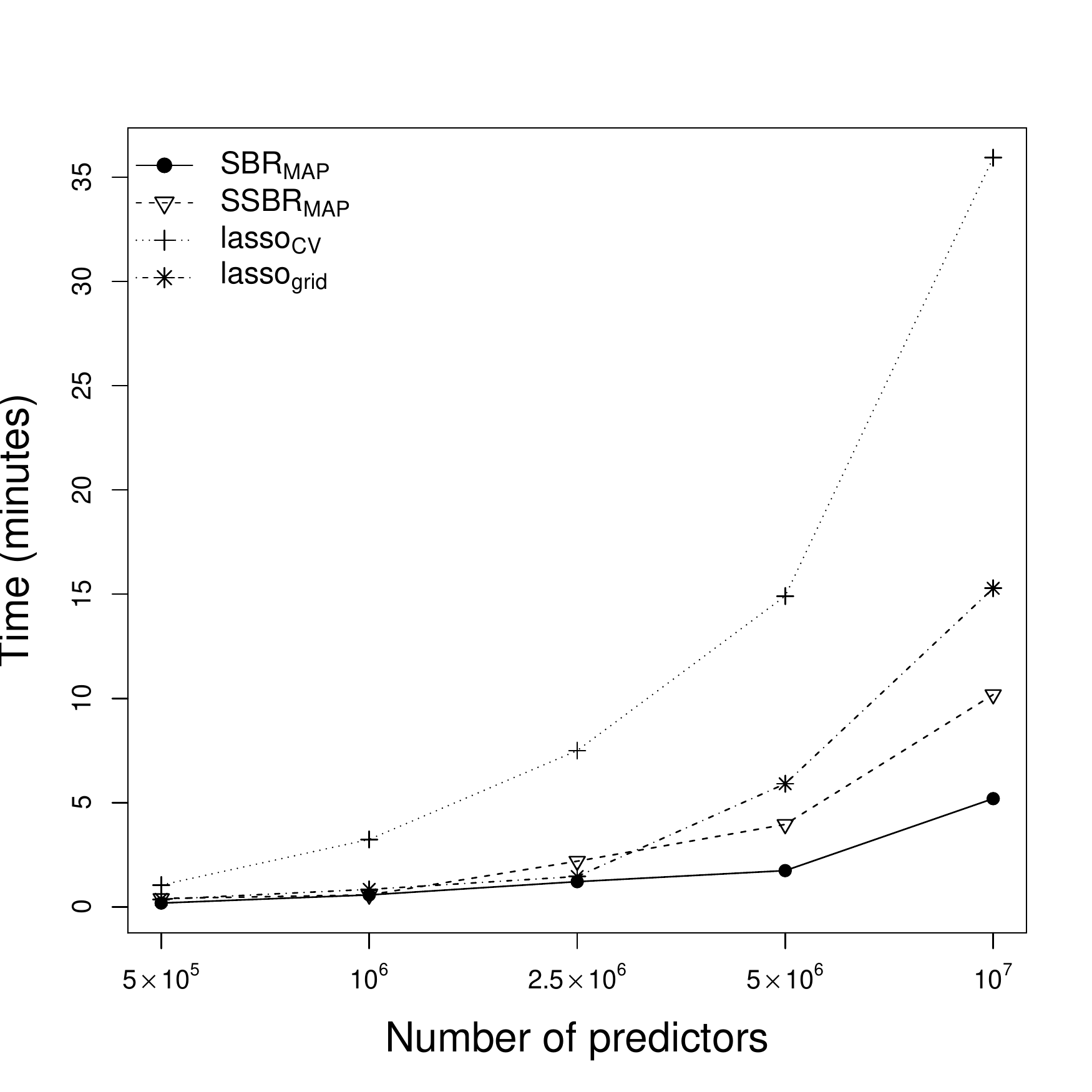}
	\caption{Simulation study, computational efficiency. Average runtime (five runs for one simulated dataset) required for SBR$_{\PM}$, SSBR$_{\PM}$, lasso$_{\CV}$ and lasso$_{\mathrm{grid}}$, for increasing $p$.}
	\label{time}
\end{figure}

Figure \ref{time} shows the average runtimes (in minutes) over five fits on a single simulated dataset. SBR and sparse SBR are considerably faster than lasso$_{\CV}$ with the gap increasing with $p$
and this is also the case in comparison to the simple lasso$_{\mathrm{grid}}$ approach described above. 
We note that by adding random noise variables as described above we in a way ``favour'' the lasso implementation in \verb|glmnet|, as the screening rules that are used by default can relatively easily  exclude these covariates.
SBR is the fastest method, with the average runtime for $p =10^7$ being approximately 5 minutes, net of all steps. Running sparse SBR with suitable parallel block-matrix calculations makes this method also very fast; average runtime was approximately 10 minutes for $p=10^7$. We note that our current implementations of the methods are certainly not optimal  in terms of computational efficiency.
We note also that our methods  can fully utilize available cores for parallel computation, hence they should continue to gain in runtime if more cores are available even with no increase in clock speed.

\section{Alzheimer's disease case study}
\label{ADNI}

\subsection{Data}
The data we consider are from  the Alzheimer's Disease Neuroimaging Initiative (ADNI) \citep{ADNI1},
a large scale longitudinal  AD study involving multiple data modalities.
The specific subset of the ADNI data we use were previously used in 
a DREAM challenge \citep{adni2} and consists of  
$n \! = \! 759$ baseline and 24 month follow-up records. Similarly to the DREAM challenge and follow-up work  \citep[e.g.,][]{Dondelinger_Mukherjee} we consider as response the scores from a cognitive function test called the mini mental state examination (MMSE); in particular, the difference in MMSE   between 24 month follow-up and baseline.


We consider three data sources: (1) clinical (CL)  data consisting of $p_{\mathrm{CL}}=12$ features (including, among others, diagnosis at baseline, Apolipoprotein E status, gender, age, years of education); (2) structural magnetic resonance imaging (MRI) data consisting of $p_{\mathrm{MRI}}=929$ features; and (3) genetic data in the form of SNP data, consisting of $p_{\mathrm{SNP}} \approx 7.3 {\times} 10^6$ features (this is the number of SNPs available after excluding those with zero variance across subjects and those with more than 10\% missing entries).
We apply the proposed methods to these data, treating the three data types (1)-(3) as sources.

The aim is to consider  a real-world application with data sources of widely differing dimensionality and to investigate whether adding the complex MRI and genetic data to the clinical covariates can improve predictive ability.
We emphasize that the goals of the present paper are mainly methodological
and that the results we present at this stage should be regarded as illustrative of the capabilities of the methods  rather than as candidate AD predictors for practical use.

\subsection{Results}
\begin{figure}[h]
	\centering{}
	\includegraphics[scale=0.5]{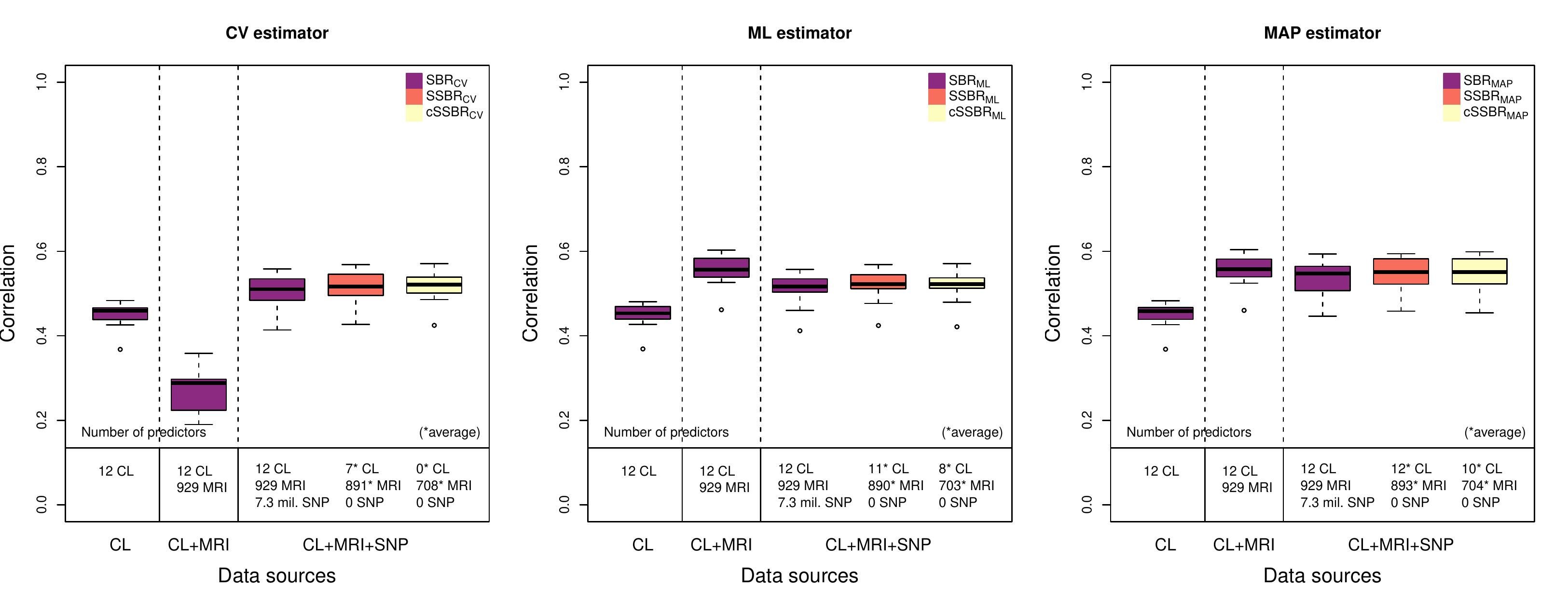}
	\caption{Alzheimer's disease case study, predictive performance. Correlations between predictions and held-out test data from 10 random splits with $n_{\mathrm{train}}=500$ and $n_{\mathrm{test}}=259$ under SBR, SSBR and cSSBR using the CV (left), ML (center) and MAP (right) estimators.}   
	\label{cor_adni}
\end{figure}

Figure \ref{cor_adni} shows results using SBR with CV, ML, and MAP estimators applied to CL only, CL and MRI, and finally all of the data (CL, MRI and SNP). For the latter case we also show results using  SSBR and cSSBR with $f_n=\log(n)$. 
Predictive performance is quantified via the correlation between predicted and observed values in held-out test data. 
The boxplots show the results of 10 random train/test splits (with $n_{\mathrm{train}}=500, \, n_{\mathrm{test}}=259$ in each split) annotated with the number of variables 
with non-zero coefficients after fitting the models in each case.

Here we see that the choice of  estimator can make a  difference: the results from the CV approach are notably worse when considering the CL and MRI datasets together. In this case the CV estimator does not adjust properly, since under the ML and MAP approaches we actually see that the addition of MRI features
to clinical covariates improves predictions.
In contrast to many  studies including MRI data, here we included all available MRI features without any pre-selection.
Adding the SNP features does not increase predictive accuracy further, rather it slightly worsens performance, in line with previous work suggesting that 
genetic data is not helpful when clinical covariates are already available \citep{adni2}.
Notably, SSBR$_\PM$ yields almost the same predictive accuracy as the regressions that do not include the SNP matrix at all; 
this is because the sparsified analyses are able to set all SNP coefficients to exactly zero. More generally, the excess risk (over CL and MRI alone) is relatively small in magnitude, despite the vast number of additional covariates. This is due to the fact that the models have a separate tuning parameter for each data source and are therefore able to effectively ``switch off''  this source, while continuing to  regularize the other covariates via source-specific penalties. In addition, cSSBR$_\PM$ provides identical predictions with SSBR$_\PM$ whilst employing fewer CL and MRI features.

\begin{figure}
	\centering{}
	\includegraphics[scale=0.48]{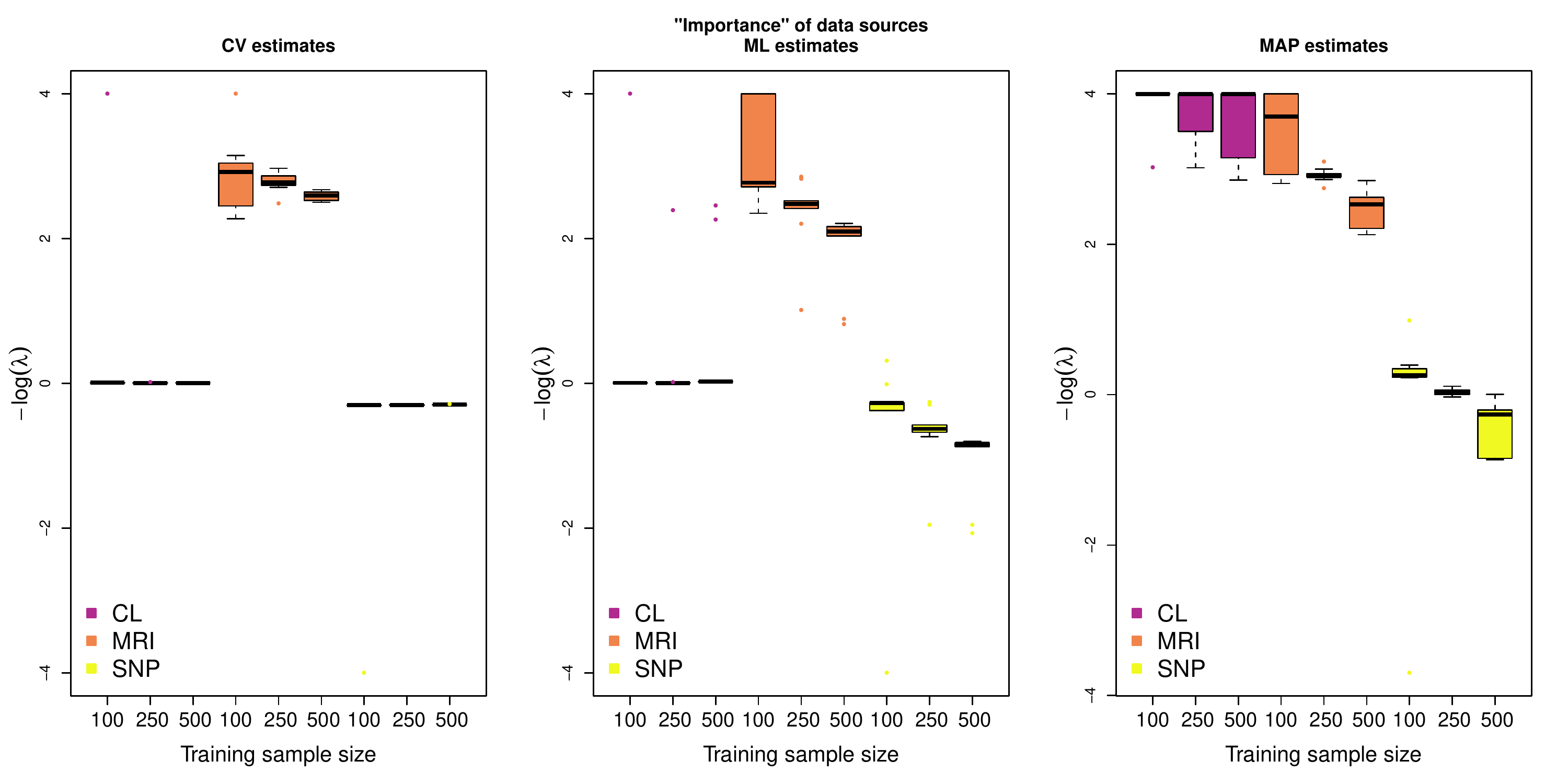}
	\caption{Alzheimer's disease  case study, estimated shrinkage levels. Boxplots show source-specific $-\log \hat{\lambda}$ values versus training sample size using the SBR methods with CV (left), ML (middle) and MAP (right) estimation of shrinkage levels.}   
	\label{lambda_adni}
\end{figure}

All penalties were estimated directly from the data and not pre-specified in any way.
To further investigate source-specific penalization, in Figure \ref{lambda_adni} we show boxplots of the source-specific $-\log \hat{\lambda}_k$ values 
(when we consider all three sources together),
for $k=\{\mathrm{CL},\mathrm{MRI},\mathrm{SNP}\}$ and $n_{\mathrm{train}}=\{100,250,500\}$.
Under the CV/ML estimators the MRI dataset is penalized less in comparison to the other two, while under the MAP estimator it is the CL dataset which appears to be the most important.
This seems to explain the slight differences between the methods (and their sparse extensions) observed in Figure \ref{cor_adni} with respect to predictive accuracy when including all sources.

A referee suggested including results from classical  ridge for the combined CL, MRI and SNP analysis. Due to computational considerations 
arising from the need to handle the large matrices (unlike in our method, classical ridge cannot decompose the problem) we consider a variant of the problem above with a random subset of half of the SNP variables. Results (again from 10 train/test splits) obtained from 5-fold CV using \verb|glmnet| indicate poor predictive performance; the resulting average correlation is 0.14 (standard deviation of 0.05). 
This result provides an empirical example of the 
benefit of  source-specific penalization.

\section{Discussion}
\label{Disc}

The aim of this paper was to introduce a   framework for high-dimensional regression  using multiple data sources  that allows efficient and fast computations in the wide data, very large $p$ setting. 
We introduced SBR, a generalized ridge-type model which can be particularly effective for prediction in the aforementioned setting. We further proposed sparse extensions (SSBR), including a general solution scalable to moderately large $p$ and a relaxed solution scalable to very large $p$. 

Concerning the three EB estimators (CV/ML/MAP), our empirical results suggest that the choice of estimator can affect predictive performance. Specifically, in the case study results under CV were worse for the combined clinical/MRI analysis, likely due to the variance of the CV risk estimate. Overall, using the MAP estimator resulted in better predictive performance both in the simulation study and in the case study; therefore, we recommend using this approach as the default option.
With respect to the sparse solutions, we mainly focused on the relaxed SSBR approaches, demonstrating that they show potential for achieving similar predictive performance to SBR but with explicit sparsity.                   
In addition, using some control for the effect of sample size on sparsity seems desirable as it can lead to enhanced sparsity with no loss in terms of prediction. 

It is worth nothing that the proposed sparsification approach based on the KL divergence is quite general in the sense that it is not restricted to the use of a ridge prior which assumes independence \textit{a-priori}. For example, one can consider a $g$-prior \citep{zellner_86} when $n>p$ or the information matrix prior \citep{gupta_ibrahim_09} when $n<p$. The latter approach, which is a combination of the $g$-prior and the ridge prior, can also be extended to incorporate multiple penalties in a manner similar to the one presented in this paper.

As a final remark, we note that this work primarily focused on predictive performance. However, we think that understanding the potential role of the methodologies presented here (and related scalable Bayesian and post-processing methods)
for variable selection will be an interesting direction for future work.


\section*{Acknowledgments}
We would like to thank the editor and anonymous referees for their constructive remarks and suggestions. Furthermore, we want to thank Rajen Shah for his insights into the equivalent representation of the general sparse solution  and Heather Cordell for input on the genetic data.\\[-0.15cm]

\noindent Data collection and sharing for the Alzheimers data application was funded by
the Alzheimers Disease Neuroimaging Initiative (ADNI) (National Institutes of
Health Grant U01 AG024904) and DOD ADNI (Department of Defense award
number  W81XWH-12-2-0012). ADNI  is funded by  the  National  Institute  on
Aging, the National Institute of Biomedical Imaging and Bioengineering, and
through generous contributions from the following:  AbbVie, Alzheimers Association;  Alzheimers Drug Discovery Foundation;  Araclon Biotech;  BioClinica, Inc.; Biogen; Bristol-Myers Squibb Company; CereSpir, Inc.; Cogstate;Eisai Inc.;
Elan Pharmaceuticals, Inc.; Eli Lilly and Company; EuroImmun; F. Hoffmann--La Roche Ltd and its affiliated company Genentech, Inc.; Fujirebio; GE Healthcare;  IXICO  Ltd.;  Janssen  Alzheimer  Immunotherapy  Research  \&  Development,  LLC.;  Johnson  \&  Johnson  Pharmaceutical  Research  \&  Development LLC.; Lumosity; Lundbeck; Merck \& Co., Inc.; Meso Scale Diagnostics, LLC.; NeuroRx Research; Neurotrack Technologies; Novartis Pharmaceuticals Corporation; Pfizer Inc.; Piramal Imaging; Servier; Takeda Pharmaceutical Company; and  Transition  Therapeutics.   The  Canadian  Institutes  of  Health  Research  is providing funds to support ADNI clinical sites in Canada.  Private sector contributions are facilitated by the Foundation for the National Institutes of Health (www.fnih.org).  The grantee organization is the Northern California Institute for Research and Education,  and the study is coordinated by the Alzheimers Therapeutic Research Institute at the University of Southern California.  ADNI data are disseminated by the Laboratory for Neuro Imaging at the University of Southern California.

\section*{Supplementary materials}

\begin{description}
	\item[Appendices:] Mathematical derivations related to Eqs. \eqref{Lambda_mode}, \eqref{Cov}, \eqref{post_sigma}, \eqref{CV}, \eqref{G_FastSol} and \eqref{relaxed} and additional results from Section 4 (see below).
	\item[R package and code: ] Package \verb|sbr| is available at \url{https://github.com/kperrakis/sbr}. (R installable files and  documentation pdf file). Reproducible code for the simulation study of Section 4 available at \url{https://github.com/kperrakis/sbr/tree/master/simulation_paper} (Rdata and R files).
\end{description}

\bibliographystyle{natbib}

\bibliography{biblio}

\newpage
\setcounter{page}{1}
\pagenumbering{roman}

\appendix
\section*{Appendices}
\renewcommand{\thesubsection}{\Alph{subsection}}
\renewcommand{\theequation}{A.\arabic{equation}} 
\setcounter{equation}{0}
\section{Computation of $\Sb$, $\bbhat$ and $b$} 
\label{AppA}

For the derivation of the posterior mode and covariance matrix in Eqs. \eqref{Lambda_mode}
and \eqref{Cov}, in the main paper, 
we use Woodbury's matrix identity \citep[p.424]{harville_97}. Specifically, we have that
\begin{align}
\Sb               & =(\La+\XT\X)^{-1} \nonumber\\
& = \La^{-1}-\La^{-1}\XT(\I+\X\La^{-1}\XT )^{-1}\X\La^{-1}\nonumber\\
& = \La^{-1}\big[\Ip-\XT(\I+\X\La^{-1}\XT)^{-1}\X\La^{-1}\big],
\label{variance}
\end{align}
so that
\begin{align}
\bbhat & =\Sb\XT\by \nonumber\\
& =\La^{-1}\big[\Ip-\XT(\I+\X\La^{-1}\XT)^{-1}\X\La^{-1}\big]\XT\by \nonumber\\
& =\La^{-1}\big[\ \XT\by-\XT(\I+\X\La^{-1}\XT)^{-1}\X\La^{-1}\XT\by\big] \nonumber\\
& =\La^{-1}\XT\big[\ \by-(\I+\X\La^{-1}\XT)^{-1}\X\La^{-1}\XT\by\big].
\label{mode}
\end{align}
For $\La$ as defined in Section 2.2 it holds that $\X\La^{-1}\XT=\sum_{k=1}^{K}\lambda_k^{-1}\Xj\XTj=\Gl$, thus, leading to \eqref{Lambda_mode}
and \eqref{Cov}.

The scale parameter $b$ of the posterior of $\sigma^2$, which is given in Eq. \eqref{post_sigma}, is derived as follows. From standard results on the conjugate normal linear model and \eqref{variance}, \eqref{mode} we have that 
\begin{align}
b & = \frac{1}{2} \big(\by^T\by -\bbhat^T\Sb^{-1}\bbhat\big) \nonumber \\
& =   \frac{1}{2} \by^T\big(\I - \X(\La+\XT\X)^{-1}\XT\big)\by.
\label{scale}
\end{align}
Now $\Hl= \X(\La+\XT\X)^{-1}\XT$ is the ``hat'' matrix in ridge regression, note however it is not idempotent ($\Hl^2 \ne \Hl$). From \eqref{variance} we have 
\begin{align*}
\Hl & = \X\Big[\La^{-1}-\La^{-1}\XT(\I+\X\La^{-1}\XT)^{-1}\X\La^{-1}\Big]\XT \\
& = \X\La^{-1}\XT-\X\La^{-1}\XT(\I+\X\La^{-1}\XT)^{-1}\X\La^{-1}\XT \\
& = \Gl-\Gl(\Ip+\Gl)^{-1}\Gl \\
& = \Big[\Gl^{-1}+\Gl^{-1}\Gl(\I+\Gl-\Gl\Gl^{-1}\Gl)^{-1}\Gl\Gl^{-1} \Big]^{-1} \mbox{     (Woodbury identity)}\\
& = (\Gl^{-1}+\I)^{-1} \\
& = \Gl(\Gl+\I)^{-1}.
\end{align*}
As a result
\begin{align}
\I-\Hl & = \I-\Gl(\Gl+\I)^{-1} \nonumber\\
& = \big[\I+\Gl(\Gl+\I-\Gl)^{-1}\big]^{-1}\mbox{     (Woodbury identity)}\nonumber\\
& = (\I+\Gl)^{-1}.
\label{res2}  
\end{align}
So from \eqref{res2}, \eqref{scale} becomes $b=\by^T(\I+\Gl)^{-1}\by / 2$ as presented in Eq. \eqref{post_sigma} in the  paper.

\renewcommand{\theequation}{B.\arabic{equation}} 
\setcounter{equation}{0}
\section{The leave-one-out CV estimator} 
\label{AppB}

The leave-one-out CV estimates are obtained via
\begin{equation}
\argmin_{\boldsymbol{\bl}}\mathrm{RSS}_{\CV}=\sum_{i=1}^{n}\big(y_i-\xti\bbhat_{\setminus i}(\boldsymbol{\bl})\big)^2.
\label{RSS}
\end{equation}
Here $\bbhat_{\setminus i}(\boldsymbol{\theta})$ is the posterior mode from the regression of $\by_{\setminus i}$ ($\by$ without the $i$-th element) on $\X_{\setminus i}$ ($\X$ without the $i$-th row). For simplicity $\bbhat_{\setminus i}\equiv\bbhat_{\setminus i}(\boldsymbol{\bl})$ henceforth. 

First, set $\A=\La+\XT\X$ and observe that
\begin{align}
(\La+\XT_{\setminus i}\X_{\setminus i})^{-1} & = (\La+\XT\X -\xii\xti)^{-1} \nonumber \\ 
& = (\A-\xii\xti)^{-1} \nonumber \\
& = \Ainv + \frac{\Ainv\xii\xti\Ainv}{1-\xti\Ainv\xii}.
\label{wood-sher}
\end{align}
For the transition from the second to the first line we used the Sherman-Morrison formula \citep[p.424]{harville_97}. 
Note that $\xti\Ainv\xii=h_{ii}$, i.e. the $i$-th element of the main diagonal of the hat matrix $\Hl=\X(\La+\XT\X)^{-1}\XT$.
Since $\bbhat_{\setminus i}=(\La+\XT_{\setminus i}\X_{\setminus i})^{-1}\XT_{\setminus i}\by_{\setminus i}=(\La+\XT_{\setminus i}\X_{\setminus i})^{-1}(\XT\by-\xii y_i)$ from the result in \eqref{wood-sher} we have
\begin{align}
\bbhat_{\setminus i} & = \Ainv(\XT\by-\xii y_i) + \frac{\Ainv\xii\xti\Ainv(\XT\by-\xii y_i)}{1-\xti\Ainv\xii} \nonumber \\
& = \bbhat -\frac{\Ainv\xii y_i(1-\xti\Ainv\xii)-\Ainv\xii\xti\Ainv(\XT\by-\xii y_i)}{1-\xti\Ainv\xii} \nonumber \\
& = \bbhat -\frac{\Ainv\xii y_i-\Ainv\xii y_i\xti\Ainv\xii-\Ainv\xii\xti\Ainv(\XT\by-\xii y_i)}{1-h_{ii}} \nonumber \\
& = \bbhat -\frac{\Ainv\xii y_i-\Ainv\xii\xti\Ainv(\xii y_i+\XT\by-\xii y_i)}{1-h_{ii}} \nonumber \\
& = \bbhat -\frac{\Ainv\xii (y_i-\xti\bbhat)}{1-h_{ii}} \nonumber \\
& = \bbhat -\frac{\Ainv\xii \varepsilon_i}{1-h_{ii}}. 
\label{estimate}
\end{align}
Pluging in \eqref{estimate} in the quantity we wish to minimize in \eqref{RSS} we get
\begin{equation*}
\mathrm{RSS}_{\CV} =\sum_{i=1}^{n}\Bigg(y_i-\xti\bbhat+\frac{\xti\Ainv\xii\varepsilon_i}{1-h_{ii}}\Bigg)^2
=\sum_{i=1}^{n}\Bigg(\varepsilon_i+\frac{h_{ii}\varepsilon_i}{1-h_{ii}}\Bigg)^2
=\sum_{i=1}^{n}\Bigg(\frac{\varepsilon_i}{1-h_{ii}}\Bigg)^2.
\end{equation*}
Switching to matrix notation we have
\begin{align}
\mathrm{RSS}_{\CV} & =(\by-\hat{\by})^T\diag(\I-\Hl)^{-2}(\by-\hat{\by}) \nonumber\\
& =\by^T(\I-\Hl)\diag(\I-\Hl)^{-2}(\I-\Hl)\by.
\label{res1}
\end{align}
The derivation up to \eqref{res1} is as in \cite{meijer} who considers the OLS case.
From \eqref{res1}  and the result in \eqref{res2} we obtain the simpler and faster solution that is used in Eq. \eqref{CV} in the paper, i.e. $\mathrm{RSS}_{\CV}=\by^T(\I+\Gl)^{-1}\big[\diag(\I+\Gl)^{-1}\big]^{-2}(\I+\Gl)^{-1}\by$.

\renewcommand{\theequation}{C.\arabic{equation}} 
\setcounter{equation}{0}
\section{Derivation of the equivalent SSBR solution} 
\label{AppC}

Initially, it is straightforward to see that for $\M=(\I+\Gl)^{-\frac{1}{2}}\X\La^{-\frac{1}{2}}$ we have
\begin{equation}
\Sb=\La^{-1}\big[\Ip-\XT(\I+\Gl)^{-1}\X\La^{-1}\big]=\La^{-\frac{1}{2}}(\Ip-\M^T\M)\La^{-\frac{1}{2}}.
\label{first}
\end{equation}
Thus, we are interested in simplifying $\Sb^{-1}=\La^{\frac{1}{2}}(\Ip-\M^T\M)^{-1}\La^{\frac{1}{2}}$ appearing in Eq. \eqref{G_Sol} that gives the SSBR solution. Let the SVD of $\M$ be given by $\M=\U\D\V^T$, where $\U\in\mathbb{R}^{n\times n}$, $\D\in\mathbb{R}^{n\times p}$ and $\V=[\V_1 ~ \V_2]$ with $\V_1\in\mathbb{R}^{p\times n}$ and $\V_2\in\mathbb{R}^{p\times (p-n)}$. Since, $\U$ and $\V$ are unitary matrices we further have that
\begin{equation}
(\Ip-\M^T\M)^{-1}=\V(\Ip-\D^2)^{-1}\V^T.
\label{second}
\end{equation}
Now observe that the upper left $n \times n$ block of $\Ip-\D^2$ is diagonal with elements the squares of the singular values, while its lower $(p-n)\times(p-n)$ block is $\boldsymbol{I}_{p-n}$ and the rest of the matrix contains only zeros. Therefore we can write $(\Ip-\D^2)^{-1}=\Ip+\widecheck{\D}$, with $\widecheck{\D}$ having non-zero entries only along the diagonal of the top left $n \times n$ block; these are given by $\tilde{d}_i=d_i^2/(1-d_i^2)$ for $i=1,\cdots,n$, where $d_i$ is the $i$-th singular value of $\M$. Let us denote this block by $\tilde{\D}\in\mathbb{R}^{n\times n}$, then we have 
\begin{equation}
\V(\Ip-\D^2)^{-1}\V^T=\V(\Ip+\widecheck{\D})\V^T=\Ip+\V_1\tilde{\D}\V_1^T.
\label{third}
\end{equation}
From \eqref{first}, \eqref{second} and \eqref{third} we get
$
\Sb^{-1}=\La+\La^{\frac{1}{2}}\V_1\tilde{\D}\V_1^T\La^{\frac{1}{2}}
$   
and it becomes clear that
\begin{equation*}
(\bbhat-\g)^T\Sb^{-1}(\bbhat-\g)=(\bbhat-\g)^T\La(\bbhat-\g)+(\bbhat-\g)^T\La^{\frac{1}{2}}\V_1\tilde{\D}\V_1^T\La^{\frac{1}{2}}(\bbhat-\g)
\end{equation*}
which explains the equivalence between Eqs. \eqref{G_Sol} and \eqref{G_FastSol} in the main paper.
\renewcommand{\theequation}{D.\arabic{equation}}
\setcounter{equation}{0}
\setcounter{figure}{0}
\section{The relaxed SSBR solution} 
\label{AppD}

Let $f(\gamma_j) = \frac{c_{n,\bl}}{2}(\hat{\beta}_j-\gamma_j)^2v_{i}^{-1}+\alpha|\gamma_j|$, with $\hat{\beta_j}\ne 0$, and let $\hat{\gamma}_j=\argmin_{\gamma_j}f(\gamma_j)$ for $j=1,\dots,p$.
It is straightforward to see that
\begin{align}
\mbox{For} & \quad \hat{\beta}_j>0: \quad \forall \,\, \gamma_j\ge 0, \quad f(\gamma_j)\le f(-\gamma_j) \Rightarrow  \hat{\gamma}_j\ge 0 & \label{C1}\\
\mbox{For} & \quad \hat{\beta}_j<0: \quad \forall \,\, \gamma_j\le 0, \quad f(\gamma_j)\le f(-\gamma_j) \Rightarrow  \hat{\gamma}_j\le 0 &
\label{C2}
\end{align}
From \eqref{C1} and \eqref{C2} we have that $\mathrm{sign}(\hat{\gamma}_j)=\mathrm{sign}(\hat{\beta}_j)$, $\forall \,\, \hat{\gamma}_j\ne 0 $. In addition for $\gamma_j\ne 0$ $f'(\gamma_j)=(\gamma_j-\hat{\beta}_j)c_{n,\bl}v_j^{-1}+\mathrm{sign}({\gamma}_j)\alpha$, therefore
\begin{align}
f'(\hat{\gamma}_j)=0 & \Leftrightarrow \hat{\gamma}_j=\hat{\beta}_j-\mathrm{sign}(\hat{\gamma}_j)\frac{v_j}{c_{n,\bl}}\alpha \nonumber \\
& \Leftrightarrow \hat{\gamma}_j=\hat{\beta}_j-\mathrm{sign}(\hat{\beta}_j)\frac{v_j}{c_{n,\bl}}\alpha.
\label{C3}
\end{align}
Thus, when $\hat{\beta}_j>0$ from \eqref{C1} and \eqref{C3} we have that 
\begin{equation*}
\hat{\gamma}_j=\begin{cases}
\hat{\beta}_j-\frac{v_j}{c_{n,\bl}}\alpha &, ~ \mbox{if} ~ \hat{\beta}_j > \frac{v_j}{c_{n,\bl}}\alpha, \\
0 &, ~ \mbox{otherwise} 
\end{cases}
\end{equation*}
and when $\hat{\beta}_j<0$ from \eqref{C2} and \eqref{C3} we have that 
\begin{equation*}
\hat{\gamma}_j=\begin{cases}
\hat{\beta}_j+\frac{v_j}{c_{n,\bl}}\alpha &, ~ \mbox{if} ~ \hat{\beta}_j < \frac{v_j}{c_{n,\bl}}\alpha, \\
0 &, ~ \mbox{otherwise}. 
\end{cases}
\end{equation*}
Which concludes the proof. For $c_{n,\bl}=\frac{n}{q_{\bl}}$ we obtain the solution in Eq. \eqref{relaxed} in the main paper.

\renewcommand{\thefigure}{E.\arabic{figure}} 
\renewcommand{\theHfigure}{E.\arabic{figure}}
\setcounter{figure}{0}
\renewcommand{\thetable}{E.\arabic{table}}
\renewcommand{\theHtable}{E.\arabic{table}}  
\setcounter{table}{0}
\section{Further results from Section \ref{SIM_STUDY}} 
\label{AppE}

\begin{figure}[H]
	\centering{}\includegraphics[width=\columnwidth]{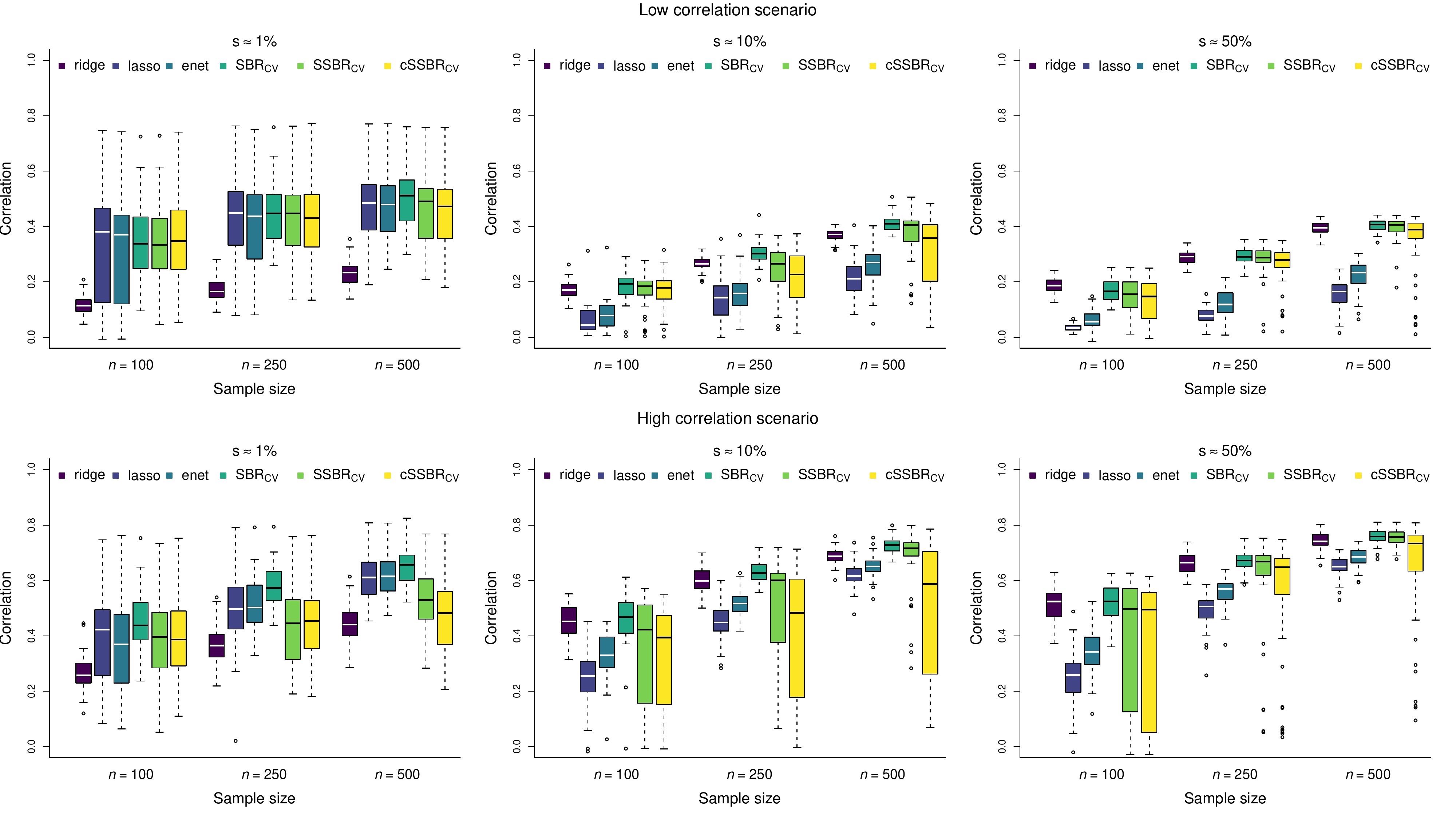}
	\caption{
		Simulation study. Comparison of ridge, lasso and elastic net to SBR, SSBR and cSSBR methods under the CV estimator for the low correlation (top panels) and high correlation (bottom panels) scenarios.
		Each panel shows correlations between predictions and held-out test data at various training sample sizes and sparsity levels as indicated. Boxplots are over 50 sampled datasets. The total dimension $p$ is over 100,000 with three (simulated) data sources (see text for details).}
	\label{cor_cv}
\end{figure}

\begin{figure}[p]
	\centering{}\includegraphics[width=\columnwidth]{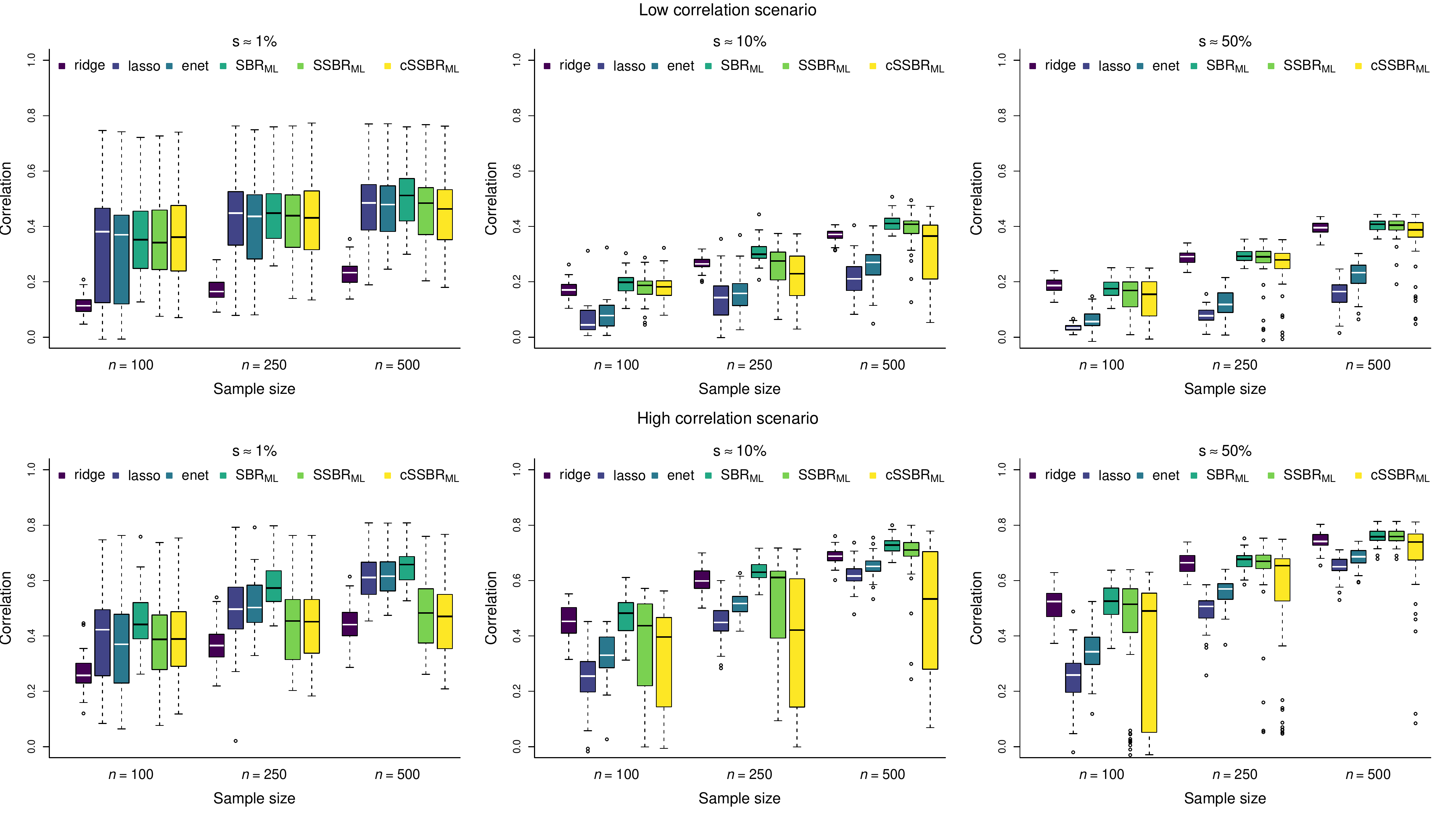}
	\caption{
		Simulation study. Comparison of ridge, lasso and elastic net to SBR, SSBR and cSSBR methods under the ML estimator for the low correlation (top panels) and high correlation (bottom panels) scenarios.
		Each panel shows correlations between predictions and held-out test data at various training sample sizes and sparsity levels as indicated. Boxplots are over 50 sampled datasets. The total dimension $p$ is over 100,000 with three (simulated) data sources (see text for details).}
	\label{cor_ml}
\end{figure}

\begin{figure}[p]
	\centering{}\includegraphics[width=\columnwidth]{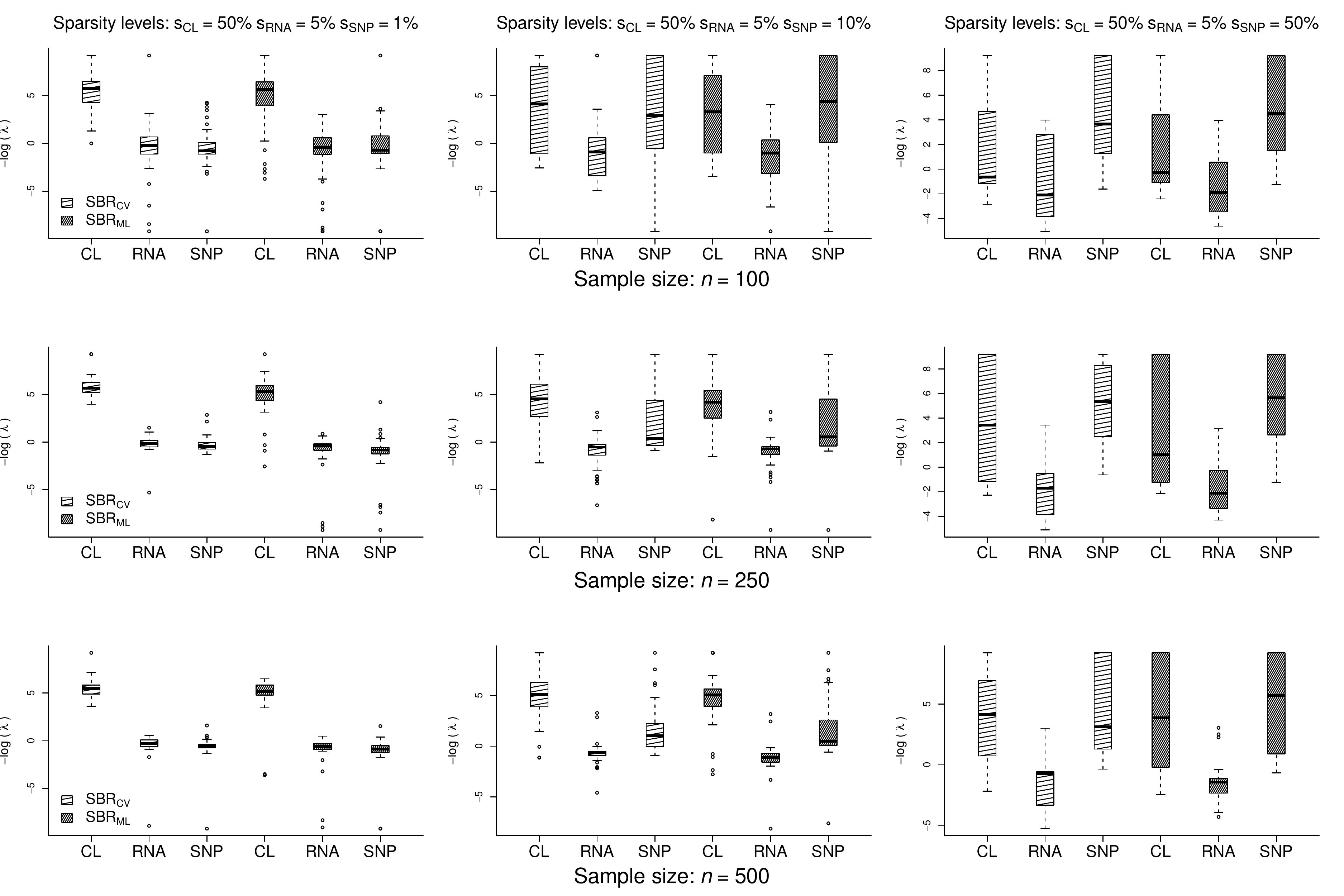}
	\caption{Simulation study, source-specific shrinkage levels. Boxplots showing source-specific $-\log\hat\lambda$ values from SBR based on the CV and ML estimators in the low-correlation simulations under various levels of sparsity for $n=100$ (top row), $n=250$ (middle row) and $n=500$ (bottom row).}
	\label{lambdas_rest}
\end{figure}

Figures \ref{cor_cv} and \ref{cor_ml} show the predictive comparisons of SBR, SSBR and cSSBR under the CV and ML estimators, respectively, with ridge, lasso and elastic net (enet) for the various sparsity levels and sample sizes under consideration. The estimated $-\log\hat{\lambda}$ under the CV and ML estimators in the low-correlation scenario are presented in Figure \ref{lambdas_rest}. Table \ref{spars1} summarizes the estimated sparsity leves from SSBR and cSSBR with CV and ML estimators, while Table \ref{spars2} shows the corresponding sparsity levels from lasso and enet. 

\begin{table}[H]
	\centering{\scalebox{0.9}{\begin{tabular}{l|c|cccccc}
				\hline
				\multirow{3}{*}{Method} & \multirow{3}{*}{Sample size} & \multicolumn{6}{c}{Simulation scenario}  \\
				&  & \multicolumn{3}{c}{Low-correlation} & \multicolumn{3}{c}{High-correlation} \\
				&                         &     $s\approx 1\%$       &      $s\approx 10\%$        &     $s\approx 50\%$     &     $s\approx 1\%$       &      $s\approx 10\%$        &     $s\approx 50\%$\\ \hline
				\multirow{3}{*}{SSBR$_\CV$}  & $n=100$   &    17\%    &     48\%       &            56\%                        &      15\%    &    45\%    &     51\% \\
				& $n=250$ &    9\%        &     40\%       & 
				70\%                        &      9\%    &    47\%        &     66\% \\
				& $n=500$ &    9\%        &     53\%       &
				69\%                        &      4\%    &    56\%        &     77\% \\ 
				\multirow{3}{*}{cSSBR$_\CV$}  & $n=100$   &    6\%    &     20\%       &            24\%                        &     4\%    &    19\%    &     23\% \\
				& $n=250$ &    2\%        &     17\%       & 
				35\%                        &      2\%    &    19\%        &     33\% \\
				& $n=500$ &    1\%        &     21\%       &
				35\%                        &      2\%    &    20\%        &     35\% \\ \hline
				\multirow{3}{*}{SSBR$_\ML$}  & $n=100$ &    19\%        &     50\%       & 
				56\%                        &      9\%    &    47\%        &     53\% \\
				& $n=250$ &    10\%        &     41\%       &
				69\%                        &      5\%    &    46\%        &     67\% \\
				& $n=500$ &    10\%        &     56\%       &
				73\%                        &      4\%    &    56\%        &     77\% \\ 
				\multirow{3}{*}{cSSBR$_\ML$}  & $n=100$ &    7\%        &     22\%       & 
				24\%                        &      3\%    &    19\%        &     23\% \\
				& $n=250$ &    3\%        &     17\%       &
				35\%                        &      1\%    &    17\%        &     34\% \\
				& $n=500$ &    2\%        &     22\%       &
				38\%                        &      0\%    &    18\%        &     38\% \\ \hline
	\end{tabular}}}
	\caption{Simulation study, induced sparsity. Average sparsity induced by the SSBR and cSSBR methods based on the CV and ML estimators over 50 repetitions of the low and high correlation simulations under varying true sparsity ($s$).}
	\label{spars1}
\end{table}

\begin{table}[H]
	\centering{\scalebox{0.9}{\begin{tabular}{l|c|cccccc}
				\hline
				\multirow{3}{*}{Method} & \multirow{3}{*}{Sample size} & \multicolumn{6}{c}{Simulation scenario}  \\
				&  & \multicolumn{3}{c}{Low-correlation} & \multicolumn{3}{c}{High-correlation} \\
				&                         &     $s\approx 1\%$       &      $s\approx 10\%$        &     $s\approx 50\%$     &     $s\approx 1\%$       &      $s\approx 10\%$        &     $s\approx 50\%$\\ \hline
				\multirow{3}{*}{lasso}  & $n=100$   &   0.02\%    &     0.008\%       &            0.016\%                        &     0.033\%    &    0.028\%    &     0.026\% \\
				& $n=250$ &    0.033\%        &     0.029\%       & 
				0.022\%                        &      0.11\%    &    0.15\%        &     0.19\% \\
				& $n=500$ &     0.099\%        &     0.093\%       &
				0.083\%                        &      0.23\%    &    0.39\%        &     0.45\% \\ \hline
				\multirow{3}{*}{elastic-net}  & $n=100$ &    0.06\%        &     0.06\%       & 
				0.09\%                        &      0.12\%    &    0.2\%        &     0.25\% \\
				& $n=250$ &    0.11\%        &     0.11\%       &
				0.13\%                        &      0.22\%    &    0.65\%        &     0.76\% \\
				& $n=500$ &    0.16\%        &     0.47\%       &
				0.54\%                        &      0.46\%    &    1.08\%        &     1.2\% \\ \hline
	\end{tabular}}}
	\caption{Simulation study, induced sparsity. Average sparsity induced by the lasso and elastic-net methods over 50 repetitions of the low and high correlation simulations under varying true sparsity ($s$).}
	\label{spars2}
\end{table}

\end{document}